\documentclass[10pt,conference]{IEEEtran}

\usepackage[T1]{fontenc}
\usepackage[utf8]{inputenc}
\usepackage{textcomp}

\usepackage[table,dvipsnames]{xcolor}
\usepackage{amsmath,amssymb,amsthm,mathtools}
\usepackage{graphicx}
\usepackage{booktabs}
\usepackage{multirow}
\usepackage{makecell}
\usepackage{array}
\usepackage{colortbl}
\usepackage{tabularx}
\usepackage{threeparttable}
\usepackage[normalem]{ulem}
\usepackage{cite}

\usepackage{algorithm}
\usepackage{algorithmic}
\usepackage{tikz}
\usetikzlibrary{shapes,arrows,positioning,calc,patterns,decorations.pathreplacing,decorations.pathmorphing,fit,backgrounds,shadows}
\usetikzlibrary{arrows.meta}
\usepackage{pgfplots}
\usepgfplotslibrary{groupplots,fillbetween}
\pgfplotsset{compat=1.17}
\usepackage{listings}
\usepackage{pifont}
\usepackage{enumitem}

\newcommand{\eqfs}{\footnotesize
  \abovedisplayskip 3pt plus 1pt minus 1pt
  \belowdisplayskip 3pt plus 1pt minus 1pt
  \abovedisplayshortskip 1pt plus 1pt
  \belowdisplayshortskip 2pt plus 1pt}
\AtBeginDocument{%
  \abovedisplayskip 4pt plus 1pt minus 1pt
  \belowdisplayskip 4pt plus 1pt minus 1pt
  \abovedisplayshortskip 1pt plus 1pt
  \belowdisplayshortskip 2pt plus 1pt}

\usepackage[hyphens]{url}
\usepackage{fancyhdr}
\usepackage[hidelinks]{hyperref}
\newcommand{\hpcayear}{2027}

\graphicspath{{figures/}}
\makeatletter
\providecommand*{\input@path}{{../}}
\makeatother

\theoremstyle{definition}
\newtheorem{definition}{Definition}
\theoremstyle{remark}

\setlist[itemize]{leftmargin=*,topsep=0pt,itemsep=0pt,parsep=0pt,partopsep=0pt}
\setlist[enumerate]{leftmargin=*,topsep=0pt,itemsep=0pt,parsep=0pt,partopsep=0pt}

\definecolor{carbongreen}{RGB}{34, 139, 34}
\definecolor{energyorange}{RGB}{255, 140, 0}
\definecolor{sloviolet}{RGB}{138, 43, 226}
\definecolor{prefillbg}{HTML}{FFF9C4}
\definecolor{thinkbg}{HTML}{C8E6C9}
\definecolor{decodebg}{HTML}{E1BEE7}
\definecolor{routerbg}{HTML}{E8F5E9}
\definecolor{freqbg}{HTML}{FFF3E0}
\definecolor{optbg}{HTML}{FFCDD2}
\definecolor{predbg}{HTML}{BBDEFB}
\definecolor{cachebg}{HTML}{E3F2FD}
\definecolor{arrowpurple}{HTML}{6A1B9A}
\definecolor{arrowgreen}{HTML}{2E7D32}
\definecolor{arrowblue}{HTML}{1565C0}
\definecolor{arroworange}{HTML}{E65100}
\definecolor{arrowred}{HTML}{C62828}
\definecolor{arrowgray}{HTML}{616161}
\definecolor{borderdark}{HTML}{212121}
\definecolor{csblue}{HTML}{0077BB}
\definecolor{csorange}{HTML}{EE7733}
\definecolor{csgreen}{HTML}{009988}
\definecolor{csred}{HTML}{CC3311}
\definecolor{cspurple}{HTML}{AA3377}
\definecolor{csgray}{HTML}{BBBBBB}

\definecolor{advA}{HTML}{1565C0} 
\definecolor{advB}{HTML}{C62828} 
\definecolor{advC}{HTML}{2E7D32} 
\definecolor{advD}{HTML}{E65100} 
\definecolor{advE}{HTML}{6A1B9A} 
\definecolor{advF}{HTML}{00695C} 
\definecolor{advG}{HTML}{AD1457} 
\definecolor{advH}{HTML}{5D4037} 
\definecolor{advI}{HTML}{283593} 
\definecolor{advJ}{HTML}{827717} 
\definecolor{advK}{HTML}{00838F} 

\newcommand{\implication}[1]{%
  \par\smallskip\noindent
  {\setlength{\fboxsep}{5pt}\setlength{\fboxrule}{0.8pt}%
  \fcolorbox{borderdark}{black!4}{%
    \parbox{\dimexpr\columnwidth-2\fboxsep-2\fboxrule\relax}{%
      \textbf{Implication:}~#1}}}%
  \par\smallskip}

\newcommand{\cmark}{\ding{51}}

\newcommand{\sys}{\textsc{PowerSlider}}

\newcommand{\para}[1]{\smallskip\noindent\textbf{#1.}}

\newcommand{\yueying}[1]{}
\newcommand{\udit}[1]{}
\newcommand{\esha}[1]{}
\newcommand{\yl}[1]{}

\newcommand{\rev}[1]{#1}
\newcommand{\grev}[1]{#1} 
\newcommand{\yrev}[1]{#1} 

\newcommand{\capfn}{\mathrm{cap}}
\newcommand{\TTFAT}{\textsf{TTFAT}}
\newcommand{\TBAT}{\textsf{TBAT}}
\newcommand{\TTLT}{\textsf{TTLT}}

\newcommand{\hpcasubmissionnumber}{1893}
\title{\sys{}: Exploiting Phase Asymmetry for LLM Serving under Demand Response}

\def\hpcacameraready{} 

\newcommand\hpcaauthors{Yueying Li$^{\dagger}$, Jiayang Chen$^{\dagger}$, Yuanfan Chen$^{\dagger}$, Leo Han$^{\dagger}$, Haoran Qiu$^{\ddagger}$, Esha Choukse$^{\ddagger}$, Rodrigo Fonseca$^{\ddagger}$, Udit Gupta$^{\S}$}
\newcommand\hpcaaffiliation{$^{\dagger}$Cornell University \quad $^{\ddagger}$Microsoft Azure Research \quad $^{\S}$Cornell Tech}
\newcommand\hpcaemail{}

\author{
  \ifdefined\hpcacameraready
    \IEEEauthorblockN{\hpcaauthors{}}
      \IEEEauthorblockA{
        \hpcaaffiliation{} \\
        \hpcaemail{}
      }
  \else
    \IEEEauthorblockN{\normalsize{HPCA \hpcayear{} Submission
      \textbf{\#\hpcasubmissionnumber{}}} \\
      \IEEEauthorblockA{
        Confidential Draft \\
        Do NOT Distribute!!
      }
    }
  \fi 
}

\fancypagestyle{camerareadyfirstpage}{%
  \fancyhead{}
  
  \fancyhead[C]{
    \ifdefined\aeopen
    \parbox[][12mm][t]{13.5cm}{\hpcayear{} IEEE International Symposium on High-Performance Computer Architecture (HPCA)}    
    \else
      \ifdefined\aereviewed
      \parbox[][12mm][t]{13.5cm}{\hpcayear{} IEEE International Symposium on High-Performance Computer Architecture (HPCA)}
      \else
      \ifdefined\aereproduced
      \parbox[][12mm][t]{13.5cm}{\hpcayear{} IEEE International Symposium on High-Performance Computer Architecture (HPCA)}
      \else
      \parbox[][0mm][t]{13.5cm}{\hpcayear{} IEEE International Symposium on High-Performance Computer Architecture (HPCA)}
    \fi 
    \fi 
    \fi 
    \ifdefined\aeopen 
      \includegraphics[width=12mm,height=12mm]{ae-badges/open-research-objects.pdf}
    \fi 
    \ifdefined\aereviewed
      \includegraphics[width=12mm,height=12mm]{ae-badges/research-objects-reviewed.pdf}
    \fi 
    \ifdefined\aereproduced
      \includegraphics[width=12mm,height=12mm]{ae-badges/results-reproduced.pdf}
    \fi
  }
  \fancyfoot[C]{}
}
\begin{document}
\maketitle

\ifdefined\hpcacameraready 
  \thispagestyle{camerareadyfirstpage}
  \pagestyle{empty}
\else
  \thispagestyle{plain}
  \pagestyle{plain}
\fi

\newcommand{\hpcaheight}{0mm}
\ifdefined\eaopen
\renewcommand{\hpcaheight}{12mm}
\fi

\thispagestyle{empty} 

\begin{abstract}
AI inference clusters are increasingly constrained by \emph{instantaneous power}, not just energy: grid operators condition new capacity on demand response, imposing time-varying power caps.
Existing LLM serving systems optimize a static energy objective or shed fixed priority tiers under load; either way, goodput collapses when the power envelope moves.
An LLM pipeline is not a uniform load: compute-bound prefill loses throughput almost linearly with GPU frequency, memory-bound answer decode sustains it down to ${\sim}0.57\times$ nominal, and reasoning's \emph{thinking} phase couples KV-cache capacity to scheduling -- so a cap should be steered to where each watt costs the least performance.
\sys{} does so with a new Flex SLO contract that turns bounded user slack into an optimization constraint, prefill--think--answer disaggregation exposing per-stage frequency and KV control, and a Karush--Kuhn--Tucker (KKT) online solver re-solving within 7.7\,ms of every cap change, backed by a consolidated fail-safe that power-gates drained instances when DVFS bottoms out on static power.
On SGLang with production traces, \sys{} sustains 78.3\% online goodput at a 30\% cap reduction versus 47.6\% for the best of five baselines ($1.64\times$), holds latency-critical tails within $1.3\times$ of nominal \grev{(baselines: $2.3$--$6\times$, up to $12\times$), and delivers 92\% mean goodput through a replayed CAISO grid-emergency day bottoming at $0.41\times$ (54\% at the trough; every baseline below 7\%)}.
\end{abstract}

\section{Introduction}
\label{sec:intro}

\begin{figure}[t]
  \centering
  \includegraphics[width=0.7\columnwidth]{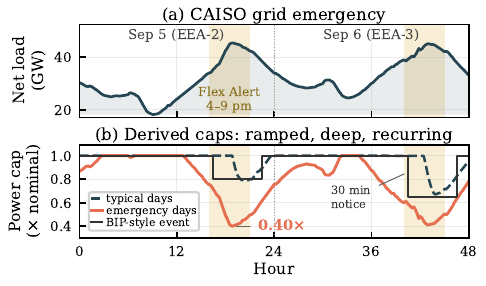}
  \vspace{-1.2em}
  \caption{Demand response is richer than a step cap. \textbf{(a)} California Independent System Operator (CAISO) net load during a grid emergencies; \textbf{(b)}~Power caps under 5 min settlement interval in high-demand and typical day~\cite{ferc2016order825}. \rev{Black steps: base interruptable program (BIP) events, dispatched 30\,min after the alert, deeper on the more severe day~\cite{pge2025bip}. EEA2/3 are grid emergency levels.}} 
  \label{fig:dr-motivation}
\end{figure}

The growth of generative AI has made data centers a first-order load on electric grids.
The U.S.\ Department of Energy projects data-center electricity consumption to grow $1.84\times$--$3.3\times$ between 2023 and 2028~\cite{doe2024datacenter,lbnl2024datacenter}, and the median wait from interconnection request to commercial operation now exceeds five years~\cite{rand2025queued}.
As a result, grid operators increasingly condition new capacity on \emph{demand response}: the ability to shed load during grid stress in exchange for faster connection and capacity credits~\cite{epri2024datacenter,bianchini2024datacenter_power,pge2025bip,iea2023dr}.
The potential is large; one study estimates that curtailing data-center load for 0.25\% of uptime could free up to 76\,GW of U.S.\ grid capacity~\cite{norris2025rethinking}.
For an AI cluster, demand response arrives as a concrete runtime constraint: a \emph{time-varying power cap} $p_{\max}(t)$ that the serving system must satisfy at every instant (Fig.~\ref{fig:dr-motivation}).
This is a fundamentally different problem from minimizing energy consumption.
\rev{An energy objective and a power cap are not interchangeable: energy is an integral that deferral can satisfy, whereas $p_{\max}(t)$ binds at every instant, and the energy-optimal operating point can itself be cap-infeasible; on A100 GPUs, energy per LLM token is minimized at an intermediate frequency (Figure~\ref{fig:freqscaling4panel}) that a deep power cap might forbid.}

\rev{Meeting a moving cap takes two ingredients.
The first is architectural:} power reduction is not uniform across an LLM serving pipeline.
Prefill phase is compute-bound and loses throughput almost linearly as GPU frequency drops.
Answer decode phase is often memory-bandwidth-bound and can run at greatly reduced frequency with little throughput loss.
Reasoning models further insert a long \emph{thinking} phase whose accumulated KV-cache state couples memory capacity to scheduling.
Hence, a demand-response event cannot be handled well by uniform power capping or by a statically profiled energy-optimal configuration; it requires a runtime that reallocates power across pipeline stages, service classes, and memory footprints as the cap changes.
This paper shows that exploiting this \emph{phase asymmetry} (in compute intensity, memory pressure, and DVFS sensitivity) is the key to sustaining goodput under deep power caps.

The second ingredient is knowing \emph{where} performance may be given up.
We observe that online inference carries significant latency slack that can be converted into power headroom.
By serving part of the traffic under a \emph{flexible} (Flex) tier that meets its base latency target most of the time but accepts bounded degradation during demand-response windows, the cluster gains room to modulate power without sacrificing any class entirely.
Commercial APIs already expose this structure: Anthropic's ``Fast'' versus ``Standard'' tiers trade up to $2.5\times$ response time for up to $6\times$ price~\cite{anthropic2026fast}, and Google's Gemini ``Flex'' tier serves the same models at reduced cost in exchange for higher, variable latency~\cite{google2025flex}.
The contract is therefore realistic; what is missing is a runtime that can honor it under a moving power envelope.

\rev{\textbf{Key insight.} A serving cluster's power flexibility depends on both system runtime and SLO contracts definition: the pipeline's phase asymmetry (where shedding watts costs little throughput) and the Flex tier's SLO tolerance (where delaying tokens costs little user harm).
We propose \sys{}, which applies this insight via a unified online optimization: it dynamically calculates both slacks to shed the watts that result in the least goodput degradation.}

Existing systems address neither half.
Energy-efficient LLM serving optimizes a \emph{static} energy objective that can be found offline~\cite{stojkovic2024dynamollm,yu2025voltanallm,zeus_nsdi,souza2023ecovisor,wiesner2021let}.
Multi-SLO serving optimizes latency and throughput across tiers but treats power as unconstrained~\cite{chen2025slosserve,zhu2025polyserve}.
Power-oversubscription work caps power with fixed priority thresholds, without per-class guarantees, time-varying caps, or support for reasoning workloads~\cite{patel2024characterizing}.
Recent power-management work spans every scope below the grid, from spatial DVFS within a GPU~\cite{kypriotis2026powerweave} to budget redistribution within a server~\cite{cho2025powersloshing} and across clusters~\cite{nam2026powergrad}, but each optimizes under a fixed envelope; none takes a moving cap as a first-class input.
Table~\ref{tab:comparison} positions \sys{} against the representative system in each category. 

To our knowledge, \sys{} is the first system to support dynamic power caps for online LLM serving with both non-reasoning and reasoning workloads.
These gaps raise three problems, which \S\ref{sec:motivation} examines with evidence: \textbf{(1)} how to expose bounded user slack without violating service guarantees, \textbf{(2)} how to search a GPU runtime configuration space efficiently, \textbf{(3)} how to handle reasoning with long, unpredictable thinking phase.

\begin{table}[t]
\centering
\scriptsize
\caption{Comparison of prior work to \sys{} } \label{tab:comparison}
\vspace{-0.4em}
\scriptsize
\setlength{\tabcolsep}{3pt}
\begin{tabular}{lcccc}
\toprule 
& Scope & Power-Aware & Reasoning & Dynamic Cap \\
\midrule
PowerWeave~\cite{kypriotis2026powerweave}  & GPU     & \cmark &  &  \\
Power Sloshing~\cite{cho2025powersloshing} & Server  & \cmark &  &  \\
\grev{\makecell[l]{PowerGrad / POLCA / Splitwise\\ DynamoLLM~\cite{nam2026powergrad,patel2024characterizing,patel2023splitwise,stojkovic2024dynamollm}}} & Cluster & \cmark &  &  \\
SLOs-Serve~\cite{chen2025slosserve}        & Cluster &  & \cmark &  \\
\midrule
\textbf{\sys{}}                             & Cluster & \cmark & \cmark & \cmark \\
\bottomrule
\end{tabular}
\end{table}

Motivated by these problems, we design \sys{}, an architecture-aware power-capping runtime for LLM inference.
Given a dynamic power cap, \sys{} jointly controls per-stage GPU allocation, stage-aware DVFS, and KV-cache memory partitioning.
A Flex SLO contract converts user-visible slack into an optimization constraint, and a prefill--think--answer (PTA) disaggregated pipeline gives the optimizer per-stage control over reasoning workloads.
Because the configuration space shifts with every cap, \sys{} replaces offline profiling with an online solver derived from the Karush--Kuhn--Tucker (KKT) conditions of a convex relaxation of the joint allocation problem.
We implement \sys{} atop SGLang~\cite{zheng2024sglang} and evaluate it on Azure LLM inference traces and reasoning workloads under swept and time-varying power caps.

In summary, this paper makes the following contributions:
\begin{itemize}[noitemsep,topsep=2pt,leftmargin=*]
\item \rev{\textbf{Characterization.} GPU LLM serving under DVFS is phase-asymmetric: compute-bound prefill loses throughput almost linearly in frequency, memory-bound answer decode holds its throughput better than prefill, and reasoning's thinking phase introduces a new memory-capacity constraint that couples KV-cache size to scheduling. \S\ref{sec:motivation} quantifies this asymmetry and its implications for power capping.}
\item \rev{\textbf{Design.} \sys{} converts that asymmetry, plus a calibrated $(\alpha_c,\rho_c)$ Flex contract, into a single runtime: PTA disaggregation exposes per-stage frequency and KV knobs, an impact metric turns contracted slack into a convex constraint, and a KKT-based solver re-solves the joint allocation problem with scalable efficiency.}
\item \rev{\textbf{Result.} On SGLang with production traces, \sys{} makes the reductions grid programs actually request (5--20\%, \S\ref{sec:grid-services}) essentially free, holding 100\% goodput for non-reasoning workloads up to a 60\% reduction, and absorbs deep caps that break every baseline: 78.3\% versus 47.6\% online goodput at a 30\% reduction under high-load reasoning ($1.64\times$), and $\ge$98\% goodput through a replayed CAISO grid-emergency day whose cap bottoms at $0.41\times$ (\S\ref{sec:eval}). The residual gap at the deepest caps is hardware, not scheduling: at $f_{\min}$, static power bounds what DVFS can shed, motivating the future power-gating direction.}
\end{itemize}

\section{Background}\label{sec:ps-background}

\subsection{Demand Response Imposes Dynamic Power Caps}\label{sec:grid-services}
\grev{Datacenters face power-modulation scenarios from millisecond-scale grid emergencies to week-long regional shortages~\cite{zhang2021flex}.}
Simultaneously, the push for sustainability drives datacenters toward renewables: wind and solar supplied a record 17\% of U.S. electricity in 2025 and roughly 90\% of new generating-capacity additions~\cite{eia2026windsolar,eia2025solaradditions}.
Their variability imposes a time-varying power envelope the datacenter must respect.
We focus on its most structured instance, \textbf{demand response} (DR): enrolled customers curtail load on request under capacity and reliability programs.
The obligation is a committed reduction against a measured baseline--commonly 5--20\% of load for C\&I participants--over events of 2-4 hours~\cite{norris2025rethinking}.
\rev{Lead time varies more, from day-ahead scheduling to the 10-minute response required of demand resources in PJM's Synchronized Reserve market~\cite{pjm2025syncreserve}; PG\&E's BIP sits near the tight end: 15--30 minutes of notice for events of up to 6 hours (at most 10 per month), with the curtailment level set by a customer-elected firm service level at least 15\% below peak demand~\cite{pge2025bip}.}

Importantly, DR imposes an \emph{instantaneous} cap $p_{\max}(t)$, not an energy budget: the serving system must stay under the envelope at every interval while maintaining SLOs.
\rev{Real grids enforce this at fine granularity: U.S. real-time markets dispatch and settle every 5 minutes under FERC Order~825~\cite{ferc2016order825}, and up to 450\,MW of flexible compute load in ERCOT already follows this 5-minute dispatch as Controllable Load Resources~\cite{ercot2025clr}.}
Throughout, we express cap severity as a \emph{power-cap reduction}: an $X\%$ reduction caps the cluster at $(100{-}X)\%$ of nominal serving power.

\subsection{LLM Serving Phases and Their Metrics}\label{sec:llm-phases}
Modern LLM inference proceeds in two classical phases.
The \emph{prefill} phase processes the input prompt, computing attention over all input tokens in parallel; it is compute-bound with high arithmetic intensity.
The \emph{decode} phase generates output tokens autoregressively; it is memory-bandwidth-bound, dominated by loading large KV-cache tensors.
This asymmetry motivated prefill--decode (PD) disaggregation~\cite{patel2023splitwise,zhong2024distserve}, which serves the two phases on separate GPUs.
\grev{Production serving systems build on this structure, from vLLM~\cite{kwon2023vllm} and Sarathi-Serve~\cite{agrawal2024sarathi} to multi-SLO schedulers such as SLOs-Serve~\cite{chen2025slosserve} and PolyServe~\cite{zhu2025polyserve}; all optimize throughput and latency, and none treats instantaneous power as a first-class constraint.}

\para{Reasoning adds a third phase}
Reasoning-capable LLMs (e.g., DeepSeek-R1~\cite{deepseekr1}, OpenAI o1/o3) generate an extended \emph{thinking} chain of internal tokens before the user-visible answer.
Thinking tokens are autoregressive like decode but with far longer sequences (often $10$--$100\times$ the answer length), making thinking phase \emph{memory-capacity}-intensive in addition to bandwidth-bound.
This introduces a new latency metrics: \emph{time-to-first-answer-token} (\TTFAT{}), the delay until the first visible output (prefill plus the entire thinking chain), which is the primary user-facing SLO for reasoning workloads; \emph{time-between-answer-tokens} (\TBAT{}), the inter-token latency of visible answer tokens; and \emph{time-to-last-token} (\TTLT{}), the end-to-end completion latency.


\section{Motivation}\label{sec:motivation}

This section establishes three observations that together define the design space.
Each observation ends with the design element it motivates; \S\ref{sec:formulation} opens with the full mapping.

\subsection{\texorpdfstring{\rev{Observation 1: Online Traffic Carries Bounded Slack that Rigid Tiers Cannot Convert into Headroom}}{Observation 1: Online Traffic Carries Bounded Slack that Rigid Tiers Cannot Convert into Headroom}}\label{sec:c1.1}

\begin{figure}[t]
    \centering
    \includegraphics[width=0.86\linewidth]{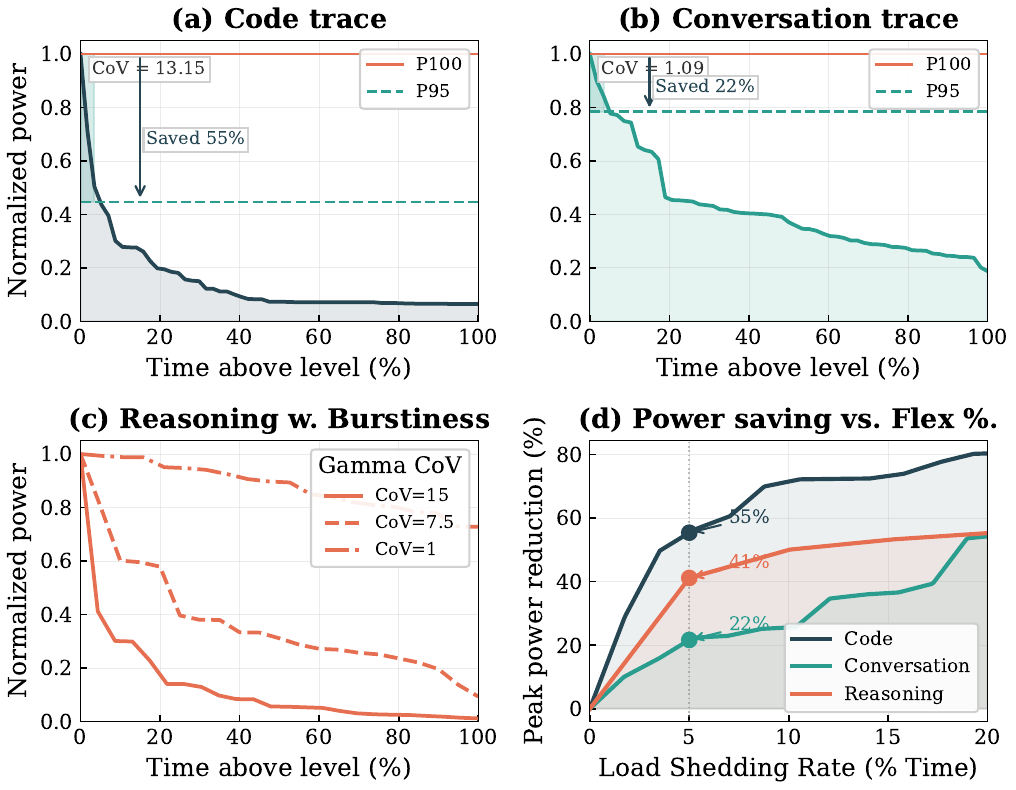}
    \vspace{-0.5em}
\caption{Load-duration trends show that peak power demand is driven by tail behavior: a small amount of service flexibility (5\%) substantially reduces required peak power provisioning. We sweep arrival burstiness (larger coefficient of variation gives more power headroom) and quantify how different load-shedding rates flatten the peak.
\rev{This burst-driven slack is the raw material \sys{} converts into demand-response headroom (O1).}}
    \label{fig:opportunity}
\end{figure}

Online LLM demand contains latency slack that can be converted into power headroom.
Figure~\ref{fig:opportunity} runs open-source non-reasoning and reasoning request traces with different arrival patterns on a cluster simulator and convert the SM / memory utilization into GPU power with profiling-based power and performance models~\cite{azure_public_dataset,magpie} .
We find that \emph{peak} power is set by rare bursts rather than sustained demand\rev{ -- a tail that survives statistical multiplexing (the traces are cluster-level aggregates), while a cap can bind even below the average draw (Fig.~\ref{fig:dr-motivation})}: allowing a small fraction of requests to experience controlled degradation, reduces the peak substantially.

\begin{figure}[t]
      \centering
      \includegraphics[width=0.78\linewidth]{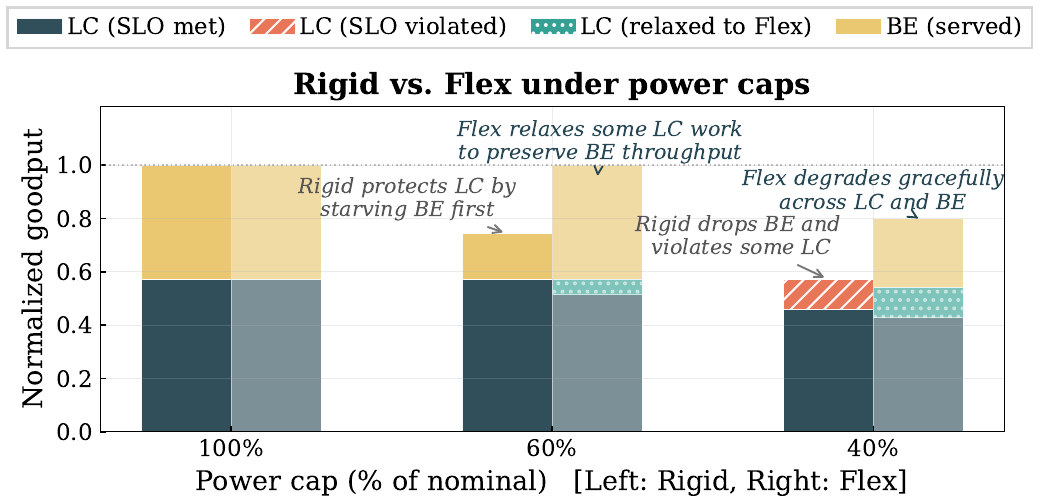}
      \vspace{-0.4em}
      \caption{\textbf{Flexible SLOs enable graceful degradation across power-cap reductions.}
At full power both paradigms deliver the same total goodput.
As the cap tightens, the rigid policy protects LC requests by rejecting BE traffic and still incurs LC violations; the flexible-SLO policy allows controlled LC degradation, preserving BE service and higher overall goodput.
\rev{Rigid tiers waste the slack of under a cap; Flex tier exists to spend it.}}
      \label{fig:rigid-vs-flex}
  \end{figure}

However, exploiting this slack requires more than relaxing SLOs. Under a rigid two-tier SLO policy (latency-critical and best-effort), the only lever when the cap tightens is shedding load: best-effort traffic is starved first, and once the cap is deep enough, latency-critical (LC) requests violate their targets as well.
Figure~\ref{fig:rigid-vs-flex} makes this concrete at 0\%, 40\%, and 60\% power-cap reductions on A100 serving a Llama-70B model: the rigid policy protects LC by rejecting best-effort (BE) traffic entirely, yet still incurs LC violations at the 60\% reduction, sharply reducing total goodput.
A \emph{Flex} tier that accepts bounded, intermittent degradation breaks this dynamic and improves goodput for both LC and BE.

The challenge, however, is not that some users can tolerate latency inflation; commercial APIs already sell such tiers~\cite{anthropic2026fast,google2025flex}.
The challenge is \emph{calibration}: enforcing and mapping a contract for SLO (e.g., at most $\alpha\times$ the base latency, for at most a $\rho$ fraction of the time) into a runtime resource-allocation decision under a changing power cap.
\implication{the system needs a Flex SLO contract that is simultaneously meaningful to users and usable as an optimization constraint (\S\ref{sec:flex-impact}).}

\subsection{Observation 2: Uniform Power Capping Wastes the Stage Asymmetry of GPU DVFS}\label{sec:c1.2}

  \begin{figure}[t]
    \centering
    \includegraphics[width=0.87\columnwidth, trim={0 0 0 28pt}, clip]{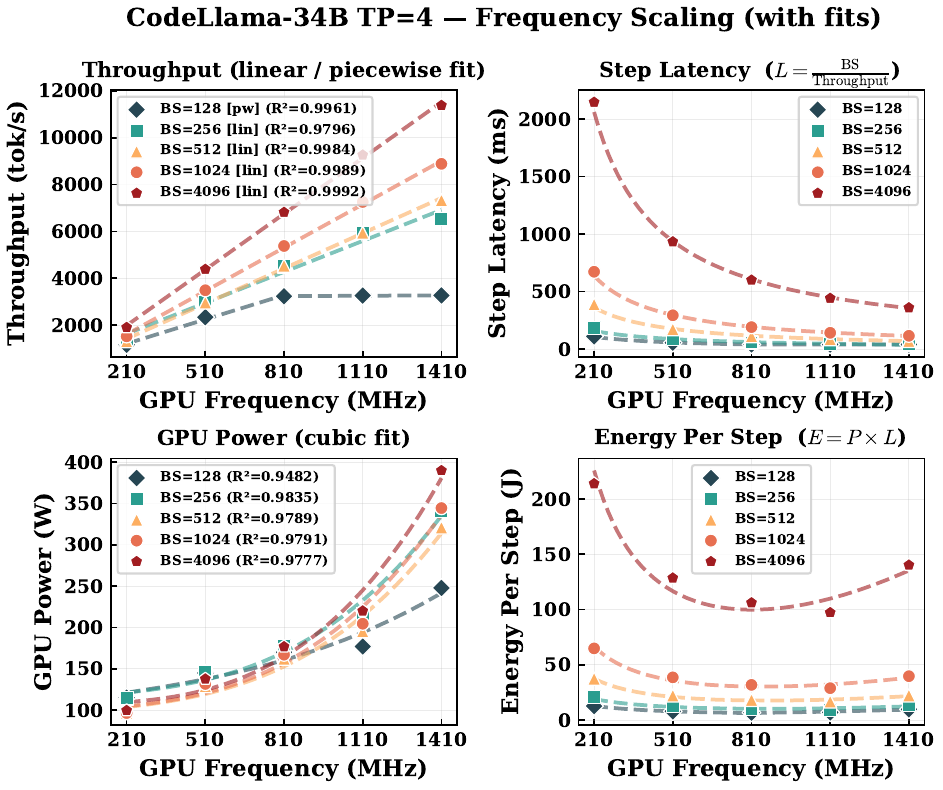}
    \vspace{-0.5em}
    \caption{Measured throughput, power, energy, and latency vs.\ GPU frequency for five batch sizes (BS) on A100-80GB.
    The frequency at which throughput saturates shifts with batch size, and BS < 128 (memory-bandwidth-bound) saturates within the DVFS range, above $810$\,MHz, saturation happens lower for lower BS; from BS>128 (compute-bound) the knee lies beyond the maximum 1{,}410\,MHz, so any frequency decrease costs a proportional throughput drop.
    \sys{} exploits this differential: it reduces batch-constrained answer-stage frequency first, shedding power at minimal performance cost.
    \rev{This stage and batch-related asymmetry is the hardware lever \sys{}'s solver pulls whenever the cap moves (O2).}}
    \label{fig:freqscaling4panel}
  \end{figure}

GPU frequency scaling does not affect all serving phases equally.
Figure~\ref{fig:freqscaling4panel} profiles CodeLlama-34B serving on A100-80GB across GPU frequencies (210-1410\,MHz) and batch sizes.
We observe two regimes.
At small batch sizes ($\le$128), throughput saturates above ${\sim}810$\,MHz ($0.57\times$ nominal); at large batch sizes ($\ge$256), it scales nearly linearly up to the maximum 1{,}410\,MHz.
A roofline analysis explains the split and why the knee shifts with different batch sizes.
The A100 can sustain ${\approx}156$ FLOP per byte of HBM traffic (312 FP16 TFLOP/s over 2.0\,TB/s); a kernel below this ridge point is limited by memory bandwidth, one above it by compute throughput.
Decode sits far below the ridge: each decode step performs roughly one multiply--accumulate per weight byte for every request in the batch, so a batch of $B$ has an arithmetic intensity of only ${\approx}B$ FLOP/byte, and charging the KV-cache reads lowers it further.
Decode throughput is therefore set by HBM bandwidth, which core DVFS does not change; the core clock must only be fast enough to keep the memory pipeline full, which on the A100 is the ${\sim}810$\,MHz knee, so frequency above the knee is wasted on decode and frequency below it starves the memory system.
Prefill is the opposite case: its multi-thousand-token GEMMs sit well above the ridge, so throughput is set by compute and falls in proportion to $f$.
Mapped onto the serving pipeline, prefill follows the compute-bound curves, answer decode at small batch the memory-bound curves, and thinking falls in between.
Concurrent work reports the same prefill/decode frequency-sensitivity split on B200-class GPUs~\cite{kypriotis2026powerweave}, indicating the asymmetry is a property of the workload rather than of one hardware generation.
Uniform power capping, which applies the same frequency to every GPU, therefore pays the \emph{worst-case} cost: it slows compute-bound prefill, where each MHz costs throughput, to save power that memory-bound decode could have given up at little performance cost.

\rev{Exploiting this asymmetry requires jointly choosing batch size, frequency, and GPU allocation per stage and SLO class, with reoptimization at every cap change (Table~\ref{tab:design-space}), since the saturation knee moves with batch size and batch formation depends on GPU allocation and KV chunk size\grev{ -- each request's per-stage KV-cache reservation, in tokens}.}

\begin{table}[t]
\centering
\scriptsize
\caption{The per-cap configuration space \sys{} must navigate.}
\label{tab:design-space}
\vspace{-0.4em}
\begin{tabular}{ll}
\toprule
\textbf{Dimension} & \textbf{Examples} \\
\midrule
Stage              & Prefill / Think / Answer \\
SLO class          & LC / Flex / BE \\
GPU allocation     & $k_i$ GPUs per stage-class \\
Frequency          & 210--1410\,MHz per stage-class \\
KV chunk size      & e.g., 512 / 1024 / 2048 tokens \\
Model / TP setting & Llama-70B, CodeLlama-34B, Qwen-32B; TP-4 \\
\bottomrule
\end{tabular}
\end{table}

This makes offline profiling impractical: with $9$ stage-class groups (3 classes $\times$ 3 stages), each choosing among ${\sim}10$ GPU-count options and ${\sim}9$ discretized frequency/chunk levels, the joint space is $(10\times 9)^{9}$ configurations, and the optimum shifts with every new cap and workload mix, so no pre-profiled grid can be dense enough.
Prior systems avoid this problem rather than solve it: DynamoLLM~\cite{stojkovic2024dynamollm} pre-profiles configurations for a single static power target and POLCA~\cite{patel2024characterizing} applies threshold-based frequency capping; neither tracks a moving cap (\S\ref{sec:ps-related}).
\implication{\rev{A dynamic power cap with various knobs across the software-hardware stack turns power allocation into an online control problem: the stage-aware optimum must be derived efficiently at event rate (\S\ref{sec:formulation-solver}) and actuated across different granularities (\S\ref{sec:ps-impl}).}}

\subsection{Observation 3: Reasoning Workloads Break PD Disaggregation and Prediction-Based Routing}\label{sec:c2}
\label{sec:challenge-prediction}

\begin{figure}[t]
  \centering
  \includegraphics[width=0.8\linewidth]{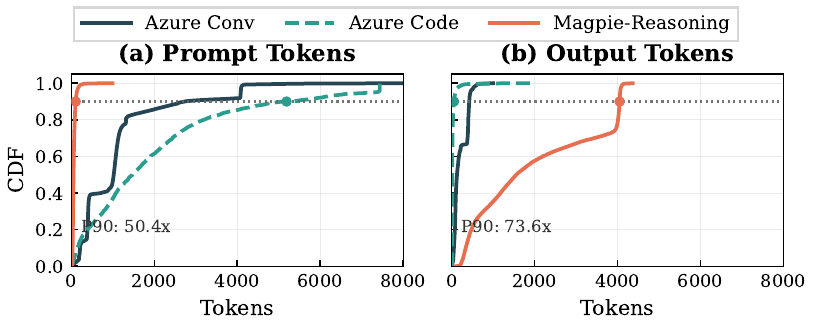}
  \vspace{-0.5em}
  \caption{Reasoning workloads have shorter prompts (left) but far longer, heavier-tailed output distributions (right), defeating the output-length prediction that prior resource-allocation and routing schemes depend on.
\rev{These tails are why \sys{} admits reasoning work by observation rather than by length prediction (O3), which matters in the near-saturation regime.}}
    \label{fig:token-distribution}
\end{figure}

\rev{A power cap pushes the cluster toward saturation, where reasoning workloads break two key mechanisms: PD disaggregation and prediction-based routing.}

\textbf{Reasoning workloads stress the memory system asymmetrically.}
Long-lived thinking chains occupy KV-cache memory for thousands to tens of thousands of tokens; when thinking and answering share the same decode workers, the accumulated thinking state caps the achievable batch size and head-of-line blocks short decode sequences. 
We can bound this pressure in bytes.
On Llama-70B, each token pins 320\,KiB of KV state, so a 4k-token thinking chain holds 1.3\,GB of HBM for its lifetime.
A TP-4 instance with ${\sim}180$\,GB of post-weight HBM capacity therefore saturates at roughly 140 such concurrent thinking-stage requests, a \emph{memory-capacity} limit on decode concurrency. 
Figure~\ref{fig:workload-sensitivity} (Appendix~\ref{sec:apx-workload-sens}) quantifies the cost: at a think-to-answer (T/A) ratio of 2:1 and 7\,req/s, P90 TBAT reaches 57\,ms under PD versus 28.3\,ms with three-stage separation (a 51\% reduction), and the gap widens monotonically with the T/A ratio (71\% at 16:1) as longer thinking chains thrash the shared KV cache.

\textbf{Prediction-based routing fails under reasoning workloads.}
Reasoning output lengths are heavy-tailed (Figure~\ref{fig:token-distribution}; the 90th percentile reaches 4050 tokens, up to $74\times$ non-reasoning workloads).
A DistBert-style length predictor as used in DynamoLLM~\cite{stojkovic2024dynamollm} misroutes 37.1\% of requests overall after fine-tuning to reasoning traces.
Among these, 26.6\% of Medium and 16.3\% of Long requests land in the Short pool, inflating its load by 25\%.
Due to the queuing effect, under high load or saturation, this inflates P99 TTFT by over an order of magnitude through routing-induced queueing (details in Appendix~\ref{sec:apx-misrouting}).
This failure mode applies broadly: any pre-sized-pool allocation fails when decode-length variance is high. Power caps exacerbate the problem by pushing the system into near-saturation state where mispredictions cascade into SLO violations.
\implication{reasoning workloads require two mechanisms: (i) stage-level isolation of thinking phase (PTA disaggregation) to prevent KV-cache thrashing, and (ii) admission control that adapts to observed output lengths rather than predicting them (\S\ref{sec:pta-admission}).}

\section{\sys{} Design}\label{sec:formulation}

\begin{figure}[t]
\centering
\includegraphics[width=0.83\columnwidth]{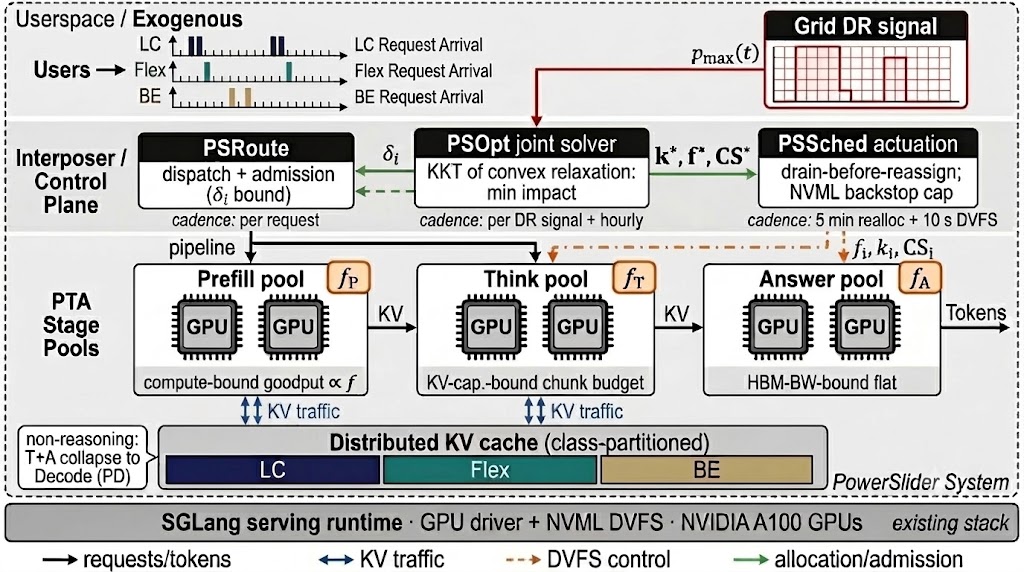}
\vspace{-0.8em}
\caption{Architecture overview of \sys{}.
Dark-headed boxes are \sys{}'s components; the gray bar is the existing stack (SGLang runtime, GPU driver, NVML DVFS).
Three SLO classes flow left-to-right through the Prefill--Think--Answer pipeline; each pool forms its own frequency domain ($f_{\mathrm{P}}, f_{\mathrm{T}}, f_{\mathrm{A}}$), and the KV cache is partitioned per class (colors match the user chips).
PSOpt turns the grid cap $p_{\max}(t)$ into per-stage allocations, frequencies, and KV budgets; PSSched actuates them; PSRoute admits and dispatches per request.}
\label{fig:design-cartoon}
\end{figure}

\sys{} is an online, scalable and dynamic architecture-aware power-capping runtime. When the grid operator lowers a cluster's power cap (e.g., by 20--40\%), \sys{} decides which stages and SLO classes to slow down, by how much, and with which control knobs.
Figure~\ref{fig:design-cartoon} summarizes the architecture.

\yrev{\sys{} comprises an optimizer, \textbf{PSOpt}, and two runtime actuators, \textbf{PSSched} and \textbf{PSRoute}; Table~\ref{tab:cadence} summarizes each component's cadence and critical-path role.
The section follows the control loop of Figure~\ref{fig:design-cartoon}: \S\ref{sec:pta-admission} presents the pipeline substrate; \S\ref{sec:flex-impact} and \S\ref{sec:formulation-solver} present PSOpt -- the contract interface that turns bounded slack into a constraint, and the solver that exploits O2's asymmetry; and \S\ref{sec:actuators} shows how PSSched and PSRoute actuate the solution.}

\begin{table}[t]
\centering
\caption{\sys{} control components}
\label{tab:cadence}
\vspace{-0.4em}
\scriptsize
\setlength{\tabcolsep}{2.5pt}
\begin{tabular}{@{}lccl@{}}
\toprule
\textbf{Component} & \textbf{Cadence} & \textbf{Crit.\ path?} & \textbf{Overhead} \\
\midrule
PSRoute (dispatch/admission) & per request & yes & virtual-queue lookup \\
\yrev{PSSched: pool reallocation}  & \yrev{5 min + 10\,s refine} & no & \yrev{drain-bounded} \\
\yrev{\quad$\hookrightarrow$ vote--commit DVFS} & \yrev{50--100\,ms} & no & \yrev{one NVML call/GPU} \\
PSOpt (KKT solve)            & DR event / hourly & no & under 100 ms \\
\bottomrule
\end{tabular}
\vspace{-0.5em}
\end{table}

\subsection{Pipeline Substrate: PTA Disaggregation}\label{sec:pta-admission}

Observation~3 (\S\ref{sec:c2}) showed that reasoning workloads defeat both PD disaggregation and prediction-based routing.
\yrev{\sys{} addresses the first with the PTA pipeline below; the second is answered by PSRoute's adaptive admission, an actuator policy (\S\ref{sec:actuators}).
PTA is the substrate the rest of the design manipulates: it defines the stage pools and the per-stage variables ($f_i$, $\mathrm{CS}_i$, $k_i$) the optimizer ranges over -- without stage isolation there is nothing stage-aware to optimize.}

\para{PTA disaggregation}
\sys{} extends PD disaggregation into the three-stage Prefill--Think--Answer pipeline, where each stage runs on a dedicated GPU pool.
Thinking tokens are generated on T-workers, isolating them from both the compute-bound prefill pipeline and the memory-capacity-bound answer decode path.
This stage isolation has two benefits.
\grev{First, each stage reserves its own chunk size ($\mathrm{CS}_T$ for think, $\mathrm{CS}_A$ for answer), whereas PD must unify the values; at typical parameters this gives PTA's think instances 44\% higher batch capacity than PD.}
Second, PTA enables independent, per-stage knobs such as frequency $f_i$: it decouples the frequency and chunk size of the two decode-like stages and exposes the stage asymmetry (Observation~2) to the solver.

\subsection{PSOpt Interface: The Flex Contract}\label{sec:flex-impact}
Central to \sys{} is serving three distinct classes of requests.
Latency-critical (LC) requests must meet strict per-workload latency targets, typically O(1--10\,s) \TTFAT{} and O(10s of ms) TBT.
Flexible (Flex) requests tolerate bounded latency degradation, formalized by the following contract.

\begin{definition}[Flex SLO Contract $(\alpha_c, \rho_c)$]
\label{def:flex-slo}
A Flex request class $c$ with base latency target $L_c$ is governed by a pair $(\alpha_c, \rho_c)$ with $\alpha_c \ge 1$ and $\rho_c \in [0,1]$. The operator may serve class-$c$ requests at a latency of up to $\alpha_c L_c$ for at most a $\rho_c$ fraction of the time during a demand-response event, and must meet $L_c$ otherwise. The contract is satisfied when
{\eqfs
\[
\Pr\big[\, L_{c}(t) > \alpha_c L_c \,\big] \;\le\; \rho_c
\]}%
over the event horizon, where $L_c(t)$ is the realized latency of class $c$ at time $t$.
\end{definition}

\noindent For example, $\alpha_c{=}2$ and $\rho_c{=}0.4$ allow up to $2\times$ latency for at most 40\% of a demand-response event.
Best-effort (BE) requests have no per-request latency target, but must finish within 24 hours and can be deferred.
\grev{This differs from threshold schemes such as POLCA~\cite{patel2024characterizing}, which cap low-priority work at fixed frequencies with no per-class guarantees or calibrated cap-to-degradation mapping.}
The $(\alpha_c, \rho_c)$ contract turns a hard SLO into a tunable probabilistic one.
Its parameters trade headroom, not safety: they bound what class-$c$ users can experience (enforced at runtime, \S\ref{sec:impl-safety}) while setting how much power headroom the operator gains.
This calibrated flexibility, rather than SLO relaxation alone, lets the system absorb power-cap reductions with limited goodput loss.

\grev{The $(\alpha_c, \rho_c)$ contract maps to a smooth impact bound: for each stage-class, there is a throughput threshold below which overload probability exceeds $\rho_c$.} $I_i(C) = (1/\bar{\lambda}_i)\smallint_{C}^{C_0} p_i(\lambda)\,(\lambda - C)\,d\lambda$ (with $C_0$ the current throughput, $p_i$ the arrival-rate distribution, and $\bar\lambda_i$ its mean).


\subsection{PSOpt Engine: Formulation and KKT Solver}\label{sec:formulation-solver}
\yrev{With the contract mapped to per-stage impact bounds and the knobs of \S\ref{sec:pta-admission} in hand, we formalize the optimization \sys{} solves at each demand-response event.}

\para{System model}
For each stage-class $i$, serving capacity depends on whether the stage is compute- or memory-bottlenecked.
Compute-limited throughput scales with GPU count $k_i$ \rev{(as $k_i^{\nu_i}$, where $\nu_i \le 1$ is the measured multi-GPU scaling exponent)} and frequency $f_i$, modulated by the compute-bound fraction $\gamma_i \in (0,1)$: prefill is highly compute-bound (high $\gamma_i$), answer decode at small batch has low $\gamma_i$, and thinking falls in between (derivation in Appendix~\ref{sec:gamma-derivation}).
Memory-limited throughput is set by how many concurrent requests fit in KV cache at the current chunk size, divided by mean service time.
Effective capacity $\capfn_i$ is the minimum of the two.

\para{Power and throughput models}
The form of the power model follows from device physics.
Dynamic CMOS power scales as $P_{\mathrm{dyn}} \propto C_{\mathrm{eff}} V^2 f$~\cite{mudge2001power,horowitz2014energy}, and on the GPU's DVFS ladder voltage rises roughly affinely with frequency ($V \approx V_0 + \beta f$)~\cite{kim2008percore,lee2011gpudvfs}, which expands to a cubic in $f$\rev{ (the expansion's $f^2$ cross-term is small over the operating range and is absorbed into the fitted coefficients)}, the form established GPU power models validate~\cite{hong2010gpupower,leng2013gpuwattch}; static leakage plus the board/HBM idle floor contribute the frequency-independent term~\cite{butts2000static}.
Per-GPU power is therefore $P(f) = c_2 f^3 + c_1 f + c_0$, with coefficients profiled per model and batch-size breakpoint (32--4096 tokens) from SGLang~\cite{zheng2024sglang} serving different model families like Llama-70B, and Qwen-32B with different TP(the characterization of Figure~\ref{fig:freqscaling4panel} is one of it); the fits achieve $R^2 > 0.95$ across profiled points.
Two architectural consequences follow.
Inverting $P(f)$ maps a per-GPU power budget to an achievable frequency floor, which lets PSOpt translate $p_{\max}(t)$ into per-stage frequency ranges. The constant term $c_0$ also limits how far DVFS can reduce power: below roughly $c_0\cdot N_{\mathrm{GPU}}$ ($\approx40\%$ of nominal on our testbed) further savings require consolidating work onto fewer GPUs, so PSOpt jointly controls $k_i$ and $f_i$.

\grev{Each hardware type needs its own cubic fit; fits are profiled offline, and since serving produces fresh (frequency, batch, latency, power) tuples each DVFS window, they can be refit online when observed latency drifts.}

\para{Joint formulation}
At each decision epoch with power cap $p_{\max}(t)$, \sys{} solves:
{\eqfs 
\begin{equation}\label{eq:dual}
\min_{\mathbf{k},\mathbf{f},\mathbf{CS}} \;
\textstyle\sum_i w_i\, I_i\!\big(C\big)
+ \kappa\|\Delta\mathbf{k}\|_1
\end{equation}}%
subject to the power cap $P \le p_{\max}(t)$, per-stage-class impact bounds $I_i \le \delta_i$, and global GPU budget $\sum_i k_i \le K$.
The objective is a weighted combination of user-visible impact, and GPU-reallocation churn\rev{; the weight $w_i$ encodes the priority of stage-class $i$ ($w_{\mathrm{LC}} \gg w_{\mathrm{Flex}} \gg w_{\mathrm{BE}}$)}.
\grev{The decision variables -- GPU allocation ($\mathbf{k}$), frequency ($\mathbf{f}$), and chunk size ($\mathbf{CS}$) -- are deeply coupled and must be optimized jointly.}

\para{Convex relaxation and KKT-guided solution}
The joint space of Table~\ref{tab:design-space} spans roughly $10^{18}$ configurations (\S\ref{sec:c1.2}), so exhaustive search, online or offline, is impractical.
Instead, \sys{} solves a \emph{convex relaxation} of the per-epoch problem: integer GPU counts $\mathbf{k}$ are relaxed to continuous values, and frequencies and chunk sizes are treated on their continuous ranges.
On the relaxation, every objective term is convex: impact satisfies $\partial^2 I_i/\partial C^2 = p_i(C)/\bar{\lambda}_i \ge 0$;
the churn term is a norm; and the budget is linear.
The KKT conditions~\cite{karush1939minima,kuhn1951nonlinear,boyd2004convex} are therefore necessary and sufficient \emph{for the relaxation}.
\rev{This is why \sys{} solves via KKT rather than a general-purpose solver: necessity and sufficiency mean the closed-form conditions \emph{are} the optimum, so each re-solve evaluates analytic gradients instead of iterating a generic convex program or integer solver, and completes in 7.7\,ms independent of cluster size (\S\ref{sec:eval-scalability}), fast enough to track every cap change online.}
At the KKT stationary point, each stage-class balances the marginal watts saved by lowering $f_i$ against the marginal impact incurred (Appendix~\ref{sec:proof-dual}).
The ranking key is the impact incurred per watt reclaimed, i.e., the impact gradient of Eq.~\eqref{eq:kkt-f} over the power gradient,

{\eqfs
\begin{equation}\label{eq:rank}
R_i \;=\; \frac{(w_i{+}\eta_i)\,\bigl|\partial I_i/\partial \capfn_i\bigr|\,\gamma_i\,k_i^{\nu_i-1}}{f_{\max}\,\bigl(3c_2 f_i^{2}+c_1\bigr)},
\end{equation}}%

and PSOpt lowers frequency in ascending order of $R_i$; this sequence is the \emph{degradation ordering}.
Memory-bound stages (low $\gamma_i$) shrink the numerator, and classes with slack impact bounds have $\eta_i{=}0$ by complementary slackness, so the solver reclaims the cheapest watts first.
This ordering restates the hardware characterization in optimization terms: $\gamma_i$ is the compute-bound fraction measured in Figure~\ref{fig:freqscaling4panel}, so in effect the solver sheds watts from the most memory-bandwidth-bound stage of the most latency-tolerant class first.

In a representative workload under a deep cap, the ordering plays out accordingly: answer decode drops onto the flat region of its throughput curve, thinking settles somewhat higher, and LC prefill stays at nominal frequency.
Uniform capping instead pulls all three stages down together, paying full throughput cost on prefill for watts that decode could have given up at little cost.
\textbf{PSOpt} evaluates these conditions directly and then \emph{projects} the relaxed solution onto discrete GPU counts, DVFS states, and chunk-size levels (full derivation in Appendix~\ref{sec:proofs}).
The projection means the deployed configuration is near-optimal for the discrete problem; the resulting degradation ordering is nonetheless the exact optimum of the relaxation, not a heuristic, and \S\ref{sec:eval-scalability} shows the full solve completes under 10 milliseconds.

\subsection{PSSched and PSRoute: Actuating the Solution}\label{sec:actuators}
\yrev{PSOpt's output is a target configuration; two actuators realize it on the cluster.}

\para{PSSched: soft stage boundaries}
\yrev{PSSched applies pool sizes and frequencies on the 5-minute loop with drain-before-reassign (no KV state is ever recomputed) and 10-second DVFS refinement between epochs, making the PTA boundary soft rather than a fixed topology.}
Pools are stage-\emph{tagged} instances within one deployment, not separate services. As arrival rates and length distributions drift, PSSched retags instances across stages, and PSRoute can borrow instances into a shared mixed pool at per-request timescale (Appendix~\ref{sec:pta-runtime}).
Thus, PD is a special case, not a separate mode: when reasoning traffic drops, the think pool drains to zero and requests follow prefill--decode; when reasoning traffic rises, the think pool regrows. 

\para{PSRoute: adaptive admission instead of prediction}
\yrev{PSRoute enforces the contract per request: class-visibility dispatch, and admission metered against the impact bound $\delta_i$, adapting to observed lengths rather than predicting them.}
\yrev{Prediction-based routing to pre-sized pools cascades under heavy-tailed reasoning lengths (\S\ref{sec:c2}); instead, PSRoute tracks a rolling average of observed per-stage token lengths as the current chunk size $\mathrm{CS}_i$.}
When outputs run shorter than expected, chunk sizes shrink and more requests are admitted; when lengths grow, concurrency is reduced to avoid KV-cache thrashing.
Within pools, PSRoute uses standard multi-SLO scheduler machinery~\cite{chen2025slosserve,zhu2025polyserve}: per-class virtual queues with priority-aware dispatch, plus pool rebalancing between PSOpt epochs (Appendix~\ref{sec:pta-runtime})\yrev{; we claim no novelty for these primitives, only for the power-aware layer above them}.

\yrev{Together the actuators close the loop across time scales -- PSOpt plans per demand-response event, PSSched moves pools in minutes, and PSSched's vote--commit DVFS daemon absorbs millisecond fluctuations; \S\ref{sec:ps-impl} details the mechanisms.}

\section{Implementation}\label{sec:ps-impl}

\sys{} builds upon SGLang~\cite{zheng2024sglang} to support a disaggregated prefill/think/answer architecture where GPU pools exchange KV cache memory via TCP or RDMA.
We structure this section around four key deployment decisions; Table~\ref{tab:cadence} outlines the execution frequency of each component and identifies which operations lie on the critical path for requests.

\para{How is GPU frequency actually set}
\sys{}'s three-stage DVFS policy changes GPU frequency at stage boundaries (e.g., from $f_{i,\text{prefill}}$ to $f_{i,\text{think}}$ when a request enters the thinking phase).
Two hardware realities shape the design.
First, NVIDIA's frequency-lock call (\textsf{nvmlDeviceSetGpuLockedClocks}) is slow relative to request timescales, taking tens to hundreds of milliseconds.
Second, stage boundaries are \emph{per-request} while DVFS is \emph{per-GPU}, creating a coordination problem: co-resident requests from different SLO classes on one GPU may want different frequencies.
\grev{We resolve both with a batched \emph{vote-commit} protocol (Algorithm~\ref{alg:vote-commit}): requests register frequency votes at stage boundaries, and every $\Delta t_{\text{dvfs}} = 50$--$100$\,ms the daemon applies the SLO-weighted argmax with a single NVML call, amortizing the transition cost across all co-resident requests.}

\begin{algorithm}[t]
\caption{Vote-commit DVFS daemon (one per GPU $g$).}
\label{alg:vote-commit}
\footnotesize
\begin{algorithmic}[1]
\STATE \textbf{state:} vote table $V_g\!: r \mapsto (f^{\star}, w)$; current frequency $f_g$
\STATE \textbf{on} request $r$ of class $c$ entering stage $\pi$ on $g$:
\STATE \quad $V_g[r] \gets (f^{\star}_{\pi,c},\, w_c)$ \hfill\COMMENT{target from PSOpt's solution}
\STATE \textbf{on} $r$ completing or migrating off $g$: \textbf{delete} $V_g[r]$
\STATE \textbf{every} $\Delta t_{\mathrm{dvfs}} = 50\text{--}100$\,ms:
\STATE \quad $f^{+} \gets \arg\max_{f \in F}\, \sum_{(f^{\star}\!,\,w) \in V_g} w \cdot b(f; f^{\star})$
\STATE \qquad \COMMENT{$b$ penalizes running a vote below its target $f^{\star}$ (SLO risk) and above it (wasted power)}
\IF{$f^{+} \neq f_g$}
  \STATE \textsc{LockClocks}$(g, f^{+})$;\; $f_g \gets f^{+}$ \hfill\COMMENT{one NVML call}
\ENDIF
\STATE \quad reassert \textsc{PowerLimit}$(g, P^{\max}_g)$ \hfill\COMMENT{hardware backstop, independent of votes (\S\ref{sec:impl-safety})}
\end{algorithmic}
\end{algorithm}

\para{How are GPUs reassigned without recomputation}
When PSSched moves a GPU between stage pools (Prefill/Think/Answer; SLO classes share these pools under PSRoute's class-visibility filters), it uses a drain-before-reassign protocol: the instance stops admitting new requests, completes or migrates the KV state of in-flight ones, and only then joins its new pool.
\grev{Draining bounds reassignment cost by the residual work of in-flight requests, at the price of reallocation latency -- hence reallocation on the slow 5-minute loop while DVFS absorbs faster fluctuations.}

\para{What does KV transfer cost}
PTA introduces one extra KV transfer per reasoning request (think $\rightarrow$ answer) than traditional PD disaggregation.
\grev{We replace SGLang-Mooncake's~\cite{qin2025mooncake} per-transfer connections with a \emph{persistent}, reference-counted session pool between worker pairs, multiplexed with sliding-window flow control against head-of-line blocking; scale-down drains rather than truncates in-flight transfers.}
An RDMA backend uses one-sided writes with the same framing, achieving 11.2\,GB/s per link versus 6.8\,GB/s for TCP on 100\,Gbps InfiniBand.
\S\ref{sec:eval-scalability} shows the transfer overhead is largely hidden by pipelining.

\para{The cap is enforced in hardware, not by the optimizer}\label{sec:impl-safety}
PSOpt's solution is an operating point, not the enforcement mechanism: PSSched always programs per-GPU NVML power limits as a backstop (the final step of Algorithm~\ref{alg:vote-commit}), so a mispredicted model or stale solve affects how much goodput survives -- never whether the cap holds beyond a bounded, sub-second transient (\S\ref{sec:eval-convergence}).
\rev{The stakes are contractual: under BIP, usage above the committed level during an event is penalized at \$6/kWh~\cite{pge2025bip}, which is why enforcement belongs in hardware rather than in the optimizer.}

\para{Frequency floor and consolidation}
The solver optimizes over $[f_{\min}, f_{\max}]$ only, the range where its fitted models are valid; if a cap is too deep to hold every active GPU even at $f_{\min}$, PSOpt reduces $k_i$ instead, consolidating load onto fewer GPUs at the next drain-bounded reallocation (until it lands, the NVML backstop holds the cap).

\para{Preventing request starvation}
LC targets are hard constraints -- BE, then Flex slack, sheds first -- and PSRoute promotes any Flex request older than its contract bound $\alpha_c L_c$ to LC-equivalent priority, so degradation ends in bounded queueing; if even LC becomes infeasible, PSOpt reports it and PSRoute sheds load by admission control rather than silently violating targets.

\section{Experimental Methodology}\label{sec:eval-setup}

\textbf{Cluster.}
\rev{We use DGX-A100 and GH200 servers as serving instances for models under different TP/EP for real-system (under 8 GPU nodes) (with SGLang + Mooncake~\cite{qin2025mooncake} KV transfer engine), scaling to 64-512 GPUs with a modified SplitwiseSim~\cite{patel2023splitwise} discrete-event simulator with profiled power and throughput model: three-stage disaggregation, KV-transfer flows, per-instance DVFS power control, and a runtime control plan implementation.
Each GPU runs at 1410\,MHz nominal with a 210\,MHz DVFS floor.
We cross-validated the simulator's testbed-profiled models (\S\ref{sec:formulation-solver}) component-wise against held-out measurements to be less than 4\% MAPE; power-fit $R^2>0.95$. Our end-to-end replay of the 8-GPU workload trace through the simulator, matches measured system goodput within 4\% and P90 \TTLT{} within 9\% over 5k requests.}

\textbf{Workload.}
\grev{We mix both reasoning and non-reasoning workload dynamically. For controlled study, we mix the Magpie-Reasoning and S1K reasoning traces~\cite{magpie,magpie_reasoning_v2_250k_r1_llama70b,simplescaling_s1k_1_1} under a bursty gamma arrival process, and Azure LLM inference traces for coding and chat with their own scaled arrival times~\cite{azure_public_dataset}.}
The default traffic mix is 30\% LC, 30\% Flex ($\alpha_c{=}3$, $\rho_c{=}0.3$), 40\% BE at saturation throughput; variations in the mix, including drift over time, are covered in Appendix~\ref{sec:apx-planned} (Figures~\ref{fig:adv-mix}). 
Workload and model details are summarized in Table~\ref{tab:workloads}.

\textbf{Demand-response scenarios.}
\grev{We sweep both static power-cap reduction levels (0--60\%) and real dynamic demand response trace for 24 hours (\S\ref{sec:eval-convergence}).} \rev{At saturation of high reasoning-mixture load, deeper caps fall below the power the cluster can shed through DVFS alone (the static-power floor of \S\ref{sec:formulation-solver}) and engage the consolidation fail-safe of \S\ref{sec:ps-impl}; the power-tracking study (Figure~\ref{fig:adv-power-tracking}) examines that regime}.

\textbf{Metrics.}
Per-class goodput (fraction of requests meeting their SLO), P90/P99 \TTFAT{} and \TTLT{}, and normalized BE throughput.
Latency is measured only for successfully completed requests in the stable window, excluding warmup and draining.

\textbf{Baselines.}
\textbf{B1} Uniform — PD-disaggregated: a single GPU frequency applied cluster-wide with overlap-KV join-shortest-queue scheduling.
\textbf{B2} POLCA~\cite{patel2024characterizing} — \grev{colocated, priority-aware DVFS: low-priority requests absorb cap reductions; no probabilistic SLO tiers or per-stage control.}
\textbf{B3} SplitWise~\cite{patel2023splitwise} — \grev{PD disaggregation with stage-aware, $\gamma$-proportional DVFS and overlap-KV JSQ scheduling, but no multi-SLO awareness.}
\textbf{B4} DynamoLLM+~\cite{stojkovic2024dynamollm} — colocated with uniform DVFS plus a multi-SLO autoscaler; assumes a static power budget and cannot reconfigure online.
\textbf{B5} SLOs-Serve+~\cite{chen2025slosserve} — multi-SLO scheduling with uniform DVFS to meet power caps.
Table~\ref{tab:baselines} summarizes these axes. Concurrent fixed-envelope systems cannot accept a moving cluster cap (\S\ref{sec:ps-related}); POLCA is the closest GPU-deployable representative.

\section{Evaluation}\label{sec:eval}

We evaluate \sys{} against the five baselines across three design axes -- architecture (PD vs.\ colocated vs.\ PTA), DVFS strategy (uniform vs.\ stage-aware vs.\ KKT-driven), and multi-SLO resource management (single-class vs.\ priority-based vs.\ Flex) -- and answer seven questions:

\textbf{Q1:} Does \sys{} maintain goodput under static power-cap reductions? (\S\ref{sec:eval-e2e})
\textbf{Q2:} Does it preserve tail latency for LC and Flex classes? (\S\ref{sec:eval-latency})
\textbf{Q3:} How sensitive is it to the Flex contract parameters $(\alpha_c,\rho_c)$? (\S\ref{sec:eval-flex})
\textbf{Q4:} Which components matter most? (\S\ref{sec:eval-ablation})
\textbf{Q5:} Can the solver and runtime adapt fast enough to a moving cap? (\S\ref{sec:eval-convergence})
\textbf{Q6:} What are the overheads, and when does PTA hurt? (\S\ref{sec:eval-scalability})
\textbf{Q7:} Does it generalize across hardware, and what does grid participation buy? (\S\ref{sec:eval-generality})

\rev{\textbf{Main result.} Figure~\ref{fig:goodput-reasoning} shows: under high-load reasoning with , \sys{} sustains $1.64\times$ more goodput than the best baseline, and Figure~\ref{fig:adv-dr-24h} shows the same system riding a real grid-emergency day at $\ge$98\% goodput, order of magnitude better.
The remaining questions explain where the gains come from (Q2--Q4), how fast the system adapts (Q5), and what it costs (Q6--Q7).} %

\subsection{Q1: End-to-End Goodput under Power-Cap Reductions}\label{sec:eval-e2e}

\begin{figure}[t]
  \centering
  \includegraphics[width=\linewidth,
                   trim={0 1.2cm 0 0}, clip]{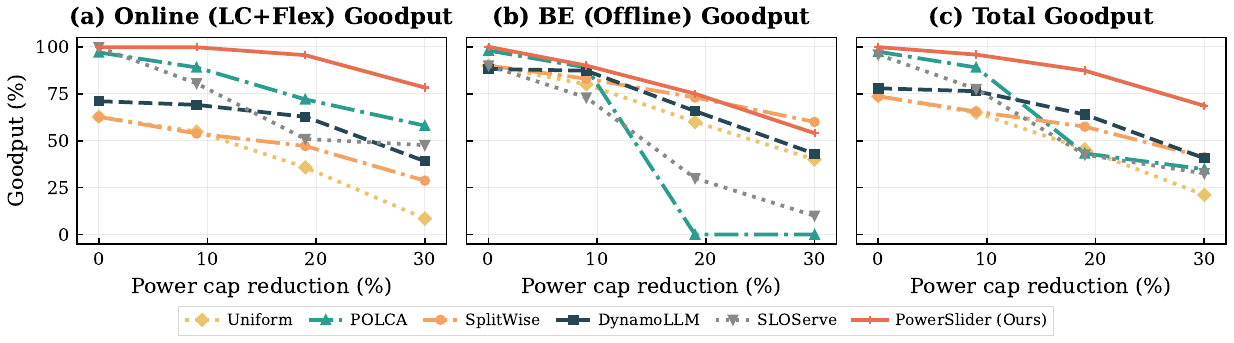}\\[-0.4ex]
  \includegraphics[width=\linewidth,
                   trim={0 0 0 0.62cm}, clip]{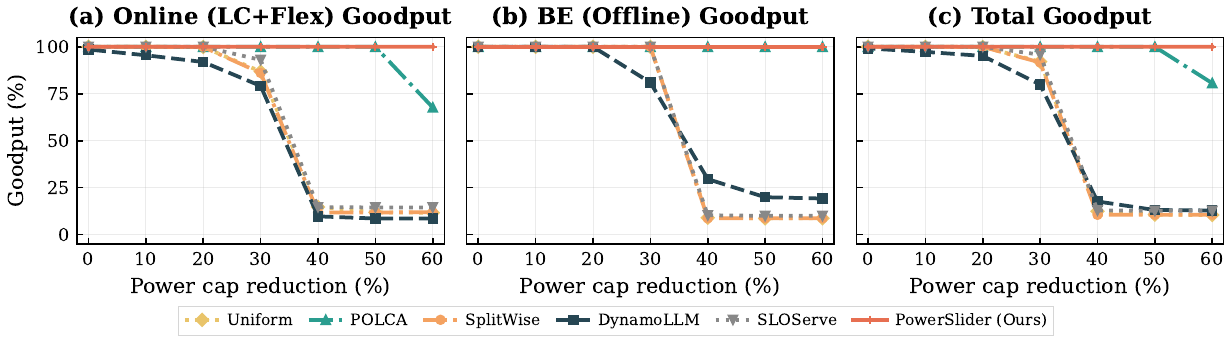}
  \vspace{-0.5em}
  \caption{Normalized online and BE goodput vs.\ power-cap reduction for \textbf{(top)} reasoning and \textbf{(bottom)} non-reasoning workloads.
\grev{\sys{} sustains $1.64\times$ the best baseline's goodput at a 30\% reduction (top) and 100\% goodput for all classes through a 60\% reduction (bottom).}
\rev{Main end-to-end result: \sys{} widens the demand-response operating envelope, making the 5--20\% reductions grid programs request (\S\ref{sec:grid-services}) essentially free.}}
  \label{fig:goodput-reasoning}\label{fig:goodput-nonreasoning}
\end{figure}

\grev{Across the sweep of different workloads, Figure~\ref{fig:goodput-reasoning} shows \sys{} preserves goodputs at power reductions where prior systems sacrifice online traffic or starve BE jobs.}

\emph{Non-reasoning (bottom).}
At QPS\,=\,14 under bursty Gamma arrivals (CoV\,=\,7.5, Azure Function Trace), \sys{} holds 100\% goodput for all three classes through a 60\% reduction; POLCA drops to 67.7\% online, and every other baseline falls below 15\% beyond a 40\% reduction.
\rev{\sys{} leverages how decode pools can drop to the $0.57\times$ frequency knee (\S\ref{sec:c1.2}), roughly halving their power at negligible throughput cost, and Flex slack absorbs the remainder -- the cap becomes free exactly when its depth fits inside the hardware asymmetry plus the contracted slack.}

\emph{Reasoning (top).}
The gap widens in the 128-GPU setting (QPS\,=\,60, CoV\,=\,13), where long thinking phases are decode-heavy enough that Uniform and SplitWise reach only 63\% online goodput even uncapped.
At a 30\% reduction, \sys{} retains 78.3\% online and 54\% BE goodput versus 47.6\% online and zero BE for the best baseline (SLOs-Serve+), a $1.64\times$ improvement.
\rev{Reasoning is intrinsically harder because the cap binds alongside a second resource: thinking's KV footprint limits decode concurrency (\S\ref{sec:c2}), and reallocating watts cannot create HBM capacity. The BE column exposes the mechanism behind the gap: baselines buy their remaining online goodput by zeroing BE, whereas \sys{} funds the cap from contracted Flex slack, so shedding stays bounded and BE survives at 54\%.}
In both regimes \sys{} shifts constrained power toward the stages and classes closest to overload, while the Flex tier keeps local bottlenecks from spreading.

A time-varying-mix study (Figure~\ref{fig:adv-mix}, Appendix~\ref{sec:apx-planned}) confirms the advantage from LC-dominant to Flex-dominant traffic: 95--100\% online goodput as the Flex share drifts 5\%$\to$80\% within one run (POLCA: 10\%).

\subsection{Q2: Tail Latency under Power-Cap Reductions}\label{sec:eval-latency}

\begin{figure}[t]
    \centering
    \includegraphics[width=0.85\columnwidth]{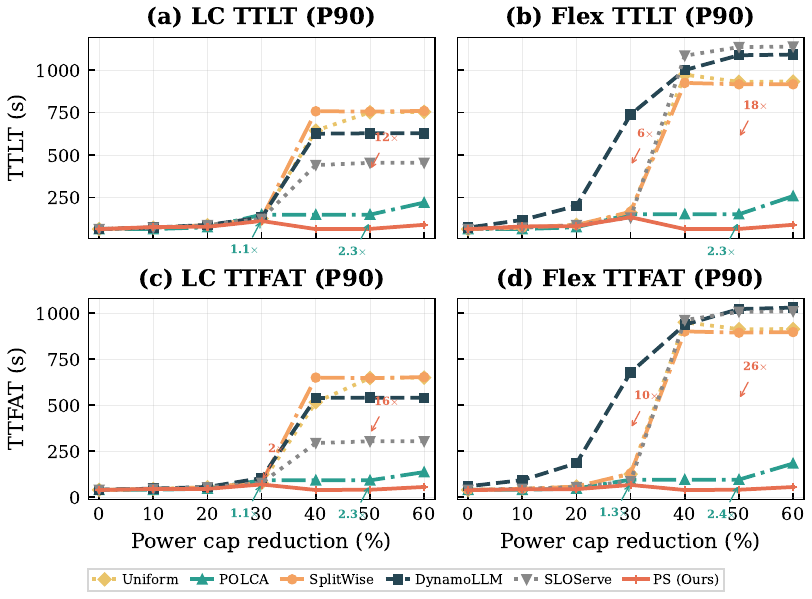}
    \vspace{-0.5em}
    \caption{P90 \TTLT{} and \TTFAT{} vs.\ power-cap reduction for the LC and Flex classes.
    \sys{} preserves near-nominal latency up to a 60\% power-cap reduction, while the best-performing baseline incurs up to $12\times$ higher LC latency and $18\times$ higher Flex latency at a 50\% reduction.
    The Flex disparity is larger because baselines do not isolate execution stages and prioritize LC at Flex's expense.
    \rev{Near-nominal tails are what keep demand-response participation compatible with latency SLOs.}}
    \label{fig:latency-vs-cap}
\end{figure}

Figure~\ref{fig:latency-vs-cap} shows P90 \TTLT{} and \TTFAT{} for LC and Flex under the bursty workload.
Even at a 60\% power-cap reduction, \sys{} keeps LC \TTLT{} at 89\,s and \TTFAT{} at 54\,s, only $1.1\times$ and $1.3\times$ above uncapped latency.
The best-performing baseline, POLCA, reaches 220\,s LC \TTLT{} and 182\,s Flex \TTFAT{}, while Uniform, SplitWise, and SLOs-Serve+ exceed 750\,s for LC and 900\,s for Flex.
The separation is driven by queueing, not by slower token generation: uniform-DVFS baselines lose compute-bound prefill throughput in proportion to frequency (\S\ref{sec:c1.2}), so under a deep cap offered load exceeds capacity and queues, and with them the tails, grow without bound, while POLCA's binary priority split protects its high tier only until the low tier is fully shed.
\sys{} keeps tails near nominal because it sheds watts where throughput is least affected and meters admission against the impact bound, so serving capacity stays ahead of admitted load and deep caps surface as bounded Flex degradation rather than queue growth\grev{, making these conclusions insensitive to the exact SLO thresholds chosen}.

\subsection{Q3: Sensitivity to the Flex Contract}\label{sec:eval-flex}

Figure~\ref{fig:flex-sensitivity} (Appendix~\ref{sec:apx-planned}) shows that \sys{} is not sensitive to the contract parameters: varying $\alpha$ or $\rho$ changes LC goodput only modestly at 20--40\% power-cap reductions.
Under deeper reductions, Flex differentiation becomes critical: at a 60\% reduction, \sys{} achieves $1.6$--$1.85\times$ the goodput of POLCA across a broad parameter range, while baselines stay near the POLCA level and Uniform drops to one-third.
Returns diminish in $\alpha$ (raising it from 2 to 10 changes little compared with 1.1 to 2): a modest contract already captures most of the power headroom.
The gain comes from the abstraction itself rather than from tuning: any reasonable $(\alpha_c,\rho_c)$ converts bounded user tolerance into a resource the solver can allocate (the impact bound $\delta_i$ of \S\ref{sec:flex-impact}), an interface the baselines lack\rev{; the \texttt{-Flex} ablation (\S\ref{sec:eval-ablation}) bounds the contract's share of the end-to-end gain} -- and since the same parameters remain an enforced ceiling on degradation (\S\ref{sec:impl-safety}), tuning moves how much headroom the operator gains, never whether the promise to users holds.

A contract-validation experiment (Figure~\ref{fig:adv-flex-cdf}) confirms $\Pr[L>\alpha_c L_c]$ stays inside the $\rho_c$ budget at every cap depth (worst 27.5\%, at a 60\% reduction); the residual mass is dominated by requests shed via admission control (\S\ref{sec:impl-safety}) rather than slow completions.

\subsection{Q4: Component Ablation}\label{sec:eval-ablation}

\begin{figure}[t]
  \centering
\includegraphics[width=0.85\linewidth]{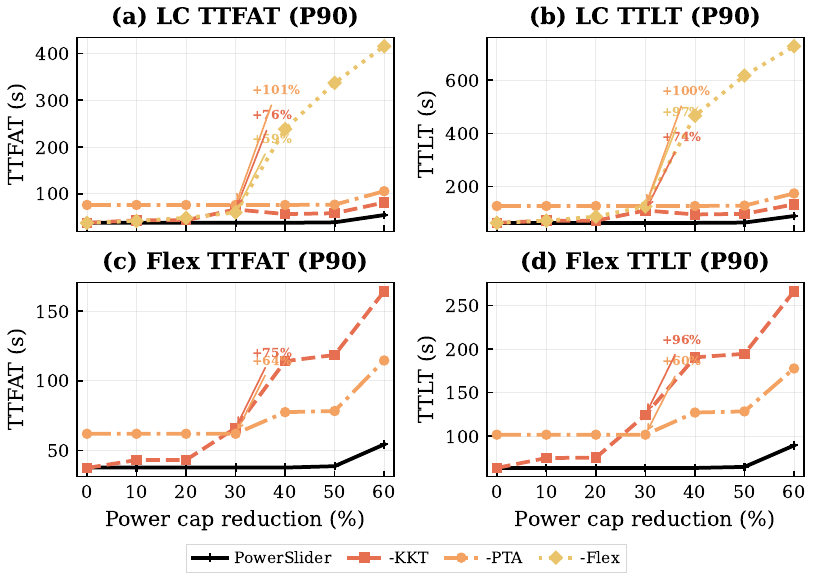}
  \vspace{-0.5em}
\caption{Ablation of \sys{} components (\texttt{-KKT}, \texttt{-PTA}, \texttt{-Flex}) for LC (top) and Flex (bottom) classes; \TTFAT{} (left) and \TTLT{} (right). The full design minimizes P90 latency across all regimes. Annotations indicate \% latency increase over \sys{} at a 30\% power-cap reduction.
\rev{Absorbing a cap needs all three mechanisms; no single knob suffices.}}
  \label{fig:ps-ablation}
\end{figure}

\textbf{\sys{}'s gains come from the interaction of all three mechanisms.}
Figure~\ref{fig:ps-ablation} shows that KKT-based power allocation, PTA disaggregation, and Flex-aware scheduling each contribute, with roles that differ by regime: at low-to-moderate power-cap reductions, accurate power allocation and stage-aware coordination alone keep latency nearly flat; as the cap tightens, PTA isolation and Flex slack become increasingly important.
\rev{The regime shift tracks which resource binds: a shallow cap is a power-allocation problem, which the solver handles alone; a deep cap becomes a memory-isolation and slack problem, where PTA and the Flex contract carry the load.}
\textbf{No single component dominates.}
Each addresses a different bottleneck -- KKT allocation distributes scarce power across groups, PTA prevents cross-stage interference, Flex converts application slack into power headroom -- and removing any one at a 30\% reduction raises LC \TTLT{} by 74--100\% and LC \TTFAT{} by 59--101\%.

\subsection{Q5: Online Adaptation and Convergence}\label{sec:eval-convergence}

\begin{figure}[t]
  \centering
  \includegraphics[width=0.77\columnwidth]{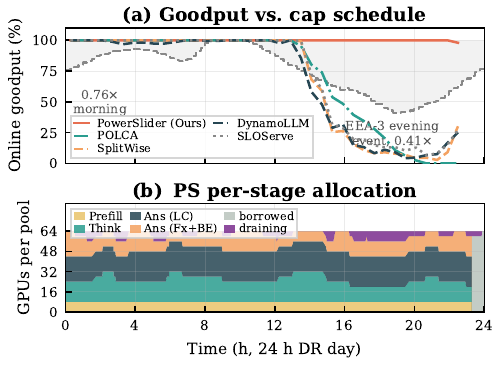}
  \vspace{-0.5em}
  \caption{\textbf{A real grid-emergency day replayed as the 24\,h DR day.} \grev{The cap schedule (gray) replays the Fig.~\ref{fig:dr-motivation} emergency day (Sep 6; trough 0.41$\times$)~\cite{gridstatus}. \textbf{(a)} \sys{} holds ${\ge}98\%$ online goodput all day; every baseline collapses to 0--6.6\% at the trough and none recovers in-trace (post-22.5\,h windows censored). \textbf{(b)} Per-stage GPU allocation under drain-before-reassign (43 reconfigurations).}}
  \label{fig:adv-dr-24h}\label{fig:adv-stage-util}
\end{figure}

\textbf{Richer DR dynamics.}
Figure~\ref{fig:adv-dr-24h} replays a real grid-emergency day rather than synthetic steps: the caps, derived from CAISO net load's EEA3 day in Fig~\ref{fig:dr-motivation}, ramp continuously, never start at full power.
\sys{} holds ${\ge}98\%$ online goodput across the entire day (worst 60\,s window: 98.0\%, at the deepest point of the trough).
The shallow morning constraint (0.76$\times$) costs the baselines almost nothing, but the evening descent breaks all of them once the cap passes ${\sim}0.8\times$: online goodput falls to 0--6.6\% near the trough (POLCA 0\%, SplitWise 2.8\%, DynamoLLM 4.2\%, SLOServe 6.6\%), and because the day ends still capped at 0.77$\times$, none re-enters the 95\%-of-baseline band within the trace.
Real cap days therefore punish the baselines harder than the stepped schedule: there is no full-power recovery window to drain their backlogs.

\textbf{Per-stage allocation.}
\grev{Figure~\ref{fig:adv-stage-util}(b) shows the allocation timeline behind this behavior. Pool moves are prefill$\leftrightarrow$think$\leftrightarrow$answer reassignments; the drain lag is visible as the band between pools, and reallocation churn strands no capacity (no pool starved).}
\rev{The cadence validates the two-time-scale split of \S\ref{sec:ps-impl}: one drain-bounded pool move per ${\sim}33$\,min suffices because the DVFS layer absorbs the 5-minute cap wiggles in between, so slow reallocation plus fast frequency control cover the grid's time scales without churn.}
\grev{As the cap steps from 90\% to 60\% of nominal on a 512-GPU cluster (Figure~\ref{fig:timeline-cap90-60}, Appendix~\ref{sec:apx-transient}), \sys{} keeps traffic stable with bounded BE latency, whereas most baselines face a brittle choice between dropping requests and instability.}

\begin{figure}[t]
  \centering
  \includegraphics[width=0.8\columnwidth]{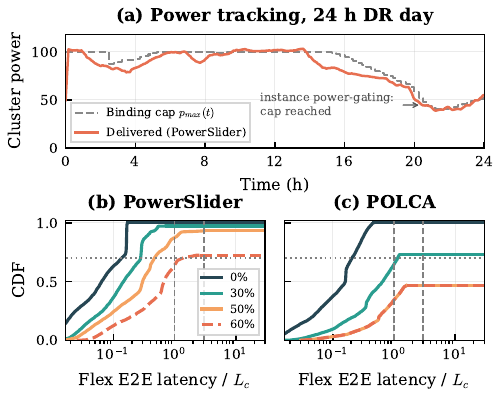}
\vspace{-0.9em}

  \caption{\textbf{Power tracking (a) and Flex contract validation (b, c) on the compressed 24\,h DR day.} (a) Delivered power on the replayed day in Fig.~\ref{fig:dr-motivation}: \sys{} stays under the binding cap throughout;
  (b) Flex E2E latency CDFs (0--60\% reductions; $L_c{=}128.9$\,s) stay inside the $(\alpha_c{=}3,\rho_c{=}0.3)$ envelope at every depth; (c) POLCA breaches it beyond 30\% power reduction}
  \label{fig:adv-power-tracking}
  \label{fig:adv-flex-cdf}
\end{figure}

\textbf{Power-tracking accuracy.}
Figure~\ref{fig:adv-power-tracking} shows per-second delivered power against the binding cap over a replayed CAISO emergency day (trough $0.40\times$): Delivered power stays under the binding cap through every event, including the trough: DVFS bottoms out on static (idle) power, and the consolidation fail-safe (\S\ref{sec:ps-impl}) power-gates drained instances to shed the remainder. The drain-bounded actuation transient exceeds the lead only on the steepest descent (4\% of event seconds at 0.5 \% of power), where the NVML backstop enforces the cap in deployment.

\rev{SLO-mix drift is equally benign: as the Flex share sweeps 5\%$\to$80\% within one run dynamically, \sys{} holds 95--100\% online goodput where POLCA collapses to 10\% (Figure~\ref{fig:adv-mix}).}

\subsection{Q6: Overheads and Failure Regimes}\label{sec:eval-scalability}

\begin{figure}[t]
  \centering
  \includegraphics[width=0.77\linewidth]{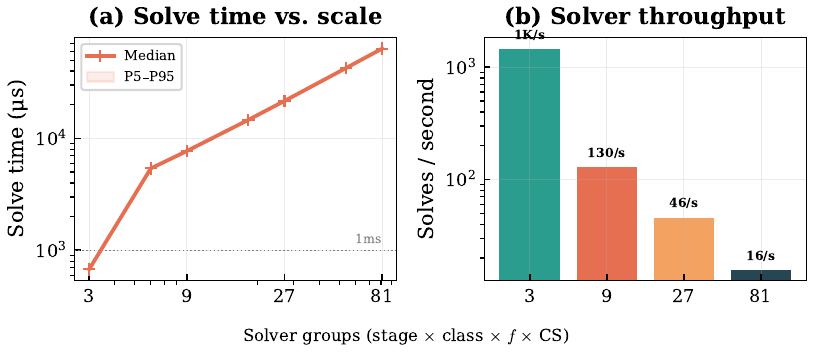}\\[2pt]
  \includegraphics[width=0.7\linewidth]{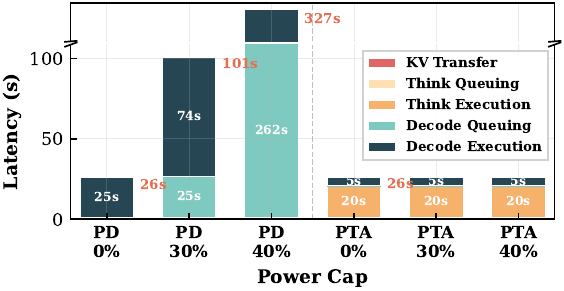}
  \vspace{-0.5em}
  \caption{\grev{\textbf{Top:} PSOpt solve time scales near-linearly with group count, staying inside one 100\,ms DVFS interval. \textbf{Bottom:} per-phase P50 latency, PD vs.\ PTA: under a 40\% cap PD develops decode queueing while PTA keeps it near zero ($12.5\times$ lower end-to-end latency).}
  \rev{Millisecond re-solves track the moving cap; stage isolation keeps it from becoming queueing collapse.}}
  \label{fig:solver-scaling}
  \label{fig:phase-breakdown}
\end{figure}

\textbf{Solver scaling.}
PSOpt assigns one decision variable per \emph{group} (a stage-class pair sharing a frequency and chunk size); the per-group GPU count enters the capacity model as an input rather than being enumerated, so solve cost is \emph{independent of cluster size and request rate}: scaling from 64 to 512 GPUs changes the values fed to the solver, not the size of the optimization.
Each solve runs 20 bisection iterations at 80 decision levels per group; in the paper's 9-group setting this takes 7.7\,ms (130 solves/s), and even at 81 groups (9 hardware types $\times$ 3 classes $\times$ 3 stages) the solve takes 63\,ms, inside a single 100\,ms DVFS interval (Figure~\ref{fig:solver-scaling}).
\rev{This is also why \sys{} needs no spatial control hierarchy (cf.\ PowerGrad's multi-level controllers~\cite{nam2026powergrad}): with solve cost independent of cluster size and per-GPU daemons acting locally on the solution, \sys{}'s hierarchy is temporal (Table~\ref{tab:cadence}), not spatial.}

\textbf{KV-transfer overhead, and when PTA hurts.}
Figure~\ref{fig:phase-breakdown} (bottom) breaks down per-phase P50 latency, including queueing and KV transfer, under long reasoning workloads (S1K, Magpie).
At nominal power, PD and PTA have similar end-to-end latency: the extra think$\to$answer handoff is hidden by pipelining, so PTA's overhead is negligible.
Under power-cap reductions the structural benefit dominates: PD develops large decode queueing as capped decode workers thrash shared KV capacity, while PTA keeps decode queueing near zero -- a $12.5\times$ lower end-to-end latency at a 40\% power-cap reduction.
\grev{PD, mixed, and full PTA are one continuum that \sys{} traverses online by retagging instances (Appendix~\ref{sec:pta-runtime}); under a drifting reasoning/non-reasoning mix, the think pool tracks the traffic share within one reallocation window. }

\subsection{Q7: Generality and the Economics of Participation}\label{sec:eval-generality}

Hardware generalization is discussed in Appendix~\ref{sec:apx-planned}: the cubic power-model form fitted on A100 also fits public H100 power curves, and porting \sys{} requires only refitting one power cubic per hardware type.


\grev{Priced at ERCOT-ERS rates (Appendix~\ref{sec:apx-planned}), the baselines' revenue loss overtakes the DR payment at 27--34\% depth; \sys{} stays net-positive throughout, serving the full load at better energy per token than uncapped (Figure~\ref{fig:adv-energy-token}(a)); no baseline reaches 90\% goodput beyond a 30\% cap.}

\subsection{Discussion: Implications for Future Hardware Design}\label{sec:eval-hw-implications}

Two hardware directions follow from our experience.
\textbf{(1) Faster, finer-grained frequency domains.}
\grev{\sys{}'s control latency is dominated by actuation, not the solver: the frequency-lock call takes tens to hundreds of milliseconds and applies to the whole GPU, forcing vote--commit batching (\S\ref{sec:ps-impl}) and limiting fast grid programs.
Per-partition domains with microsecond transitions (under 0.5\% die area~\cite{kypriotis2026powerweave}) would let power follow \emph{per-request} stage boundaries and deepen AGC-speed enrollment.}
\textbf{(2) Hardware power telemetry and enforcement.}
\grev{Grid participation requires trustworthy caps, but today's enforcement loop is firmware-mediated and opaque; on-package capping with bounded response time and per-SM/HBM attribution would enable contractual exceedance bounds and compute-granularity power management.}

\section{Related Work}\label{sec:ps-related}

\textbf{Power and latency control in datacenters.}
Pegasus~\cite{lo2014pegasus}, Heracles~\cite{lo2015heracles}, Adrenaline~\cite{hsu2015adrenaline}, PARTIES~\cite{chen2019parties}, and Caladan~\cite{Fried2020} established power and frequency as runtime actuators under latency SLOs, and Dynamo~\cite{wu2016dynamo}, Thunderbolt~\cite{li2020thunderbolt}, and Flex~\cite{zhang2021flex} manage fleet-scale budgets under QoS --- on CPUs, for inferred request-level slack; \grev{\sys{} carries the premise to GPU LLM inference, where phase-dependent boundedness shapes the actuator and the slack is \emph{contractual}.}

\textbf{Fixed-envelope power management for AI hardware.}
PowerWeave~\cite{kypriotis2026powerweave} (spatial per-partition DVFS under hard SLOs), Power Sloshing~\cite{cho2025powersloshing} (CPU$\leftrightarrow$GPU redistribution of a fixed server cap), and PowerGrad~\cite{nam2026powergrad} (hierarchical steering of a cluster budget), with earlier hierarchical capping~\cite{wu2016dynamo,li2020thunderbolt}, each divide a \emph{fixed} envelope more efficiently but stop below the grid interface: none decides which stages, service classes, and memory configurations absorb a cap reduction.
\grev{None is a \emph{distinct} moving-cap baseline either: PowerWeave takes no budget input, server-local sloshing reduces to uniform capping under a cluster-wide cut, and PowerGrad's gradient steering degenerates to proportional capping under stage disaggregation; and \sys{} supplies the grid-facing layer such controllers could enforce.}

\textbf{DVFS hardware technology.}
\grev{Silicon already offers fine, fast control: per-core regulators~\cite{kim2008percore}, integrated VRs~\cite{burton2014fivr}, digital LDOs~\cite{okuma2010dldo}, GPU DVFS prototypes~\cite{lee2011gpudvfs,sethia2014equalizer}, memory DVFS~\cite{david2011memorydvfs}, and sub-0.5\%-area per-SM domains~\cite{kypriotis2026powerweave}; shipping GPUs expose one coarse frequency domain per die.
\sys{} is built for that interface with a granularity-agnostic solver: finer domains would deepen fast grid participation.}

\textbf{Multi-class QoS and interference management.}
Paragon~\cite{delimitrou2013paragon}, Quasar~\cite{delimitrou2014quasar}, Bubble-Up/Flux~\cite{mars2011bubbleup,yang2013bubbleflux}, and their Tarcil~\cite{delimitrou2015tarcil,petrucci2015octopus,kulkarni2020cuttlesys,xu2013bobtail} deliver graceful per-class degradation for cache, cores, and memory bandwidth; \grev{POLCA~\cite{patel2024characterizing} is a binary two-class instance for GPU power, and \sys{}'s $(\alpha_c,\rho_c)$ contracts and stage-aware allocation provide the missing analogue under varying caps.}

\textbf{GPU and ML energy--performance tradeoffs.}
Zeus~\cite{zeus_nsdi} and Perseus~\cite{chung2024perseus} trace energy--time Pareto frontiers for training, the view our power-goodput frontier builds on; $\mu$-Serve, DynamoLLM, TAPAS, VoltanaLLM, and EcoServe~\cite{qiu2024muserve,stojkovic2024dynamollm,stojkovic2025tapas,yu2025voltanallm,li2025ecoserve} apply DVFS or auto-scaling to inference under a fixed SLO or static priority split\rev{, and throttLL'eM~\cite{kakolyris2025throttllem} adds predictive throttling with instance autoscaling to minimize energy under SLOs} -- none \rev{takes a power cap as input or expresses per-class degradation under a time-varying cap.}

\textbf{Datacenter demand response and carbon-aware scheduling.}
Grid-interactive datacenters~\cite{liu2011greening,lin2011dynamic,liu2014pricing,wierman2014opportunities} and carbon-aware batch systems~\cite{goiri2011greenslot,goiri2012greenhadoop,souza2023ecovisor,wiesner2021let,hanafy2023war} shift \emph{deferrable} work at job granularity; Carbon Explorer~\cite{acun2023carbon} shows that the ability to \emph{follow} a grid signal unlocks carbon-aware operation.
\grev{CarbonScaler~\cite{hanafy2024carbonscaler} allocates time-varying capacity using per-job marginal-utility curves, but latency-bound serving offers no such curve; \sys{}'s Flex contract and power--goodput frontier provide exactly that elasticity profile, while every request still meets its SLO as the cap moves.}


\section{Conclusion}\label{sec:conclusion}

\grev{Dynamic power caps turn LLM inference into an architecture-aware runtime control problem: pipeline stages differ in compute intensity, memory pressure, DVFS sensitivity, and SLO slack, so a uniform watt costs far more than a well-placed one.}
\grev{\sys{} answers with the Flex contract, PTA disaggregation, and a KKT-guided online solver: $1.64\times$ the best baseline's goodput at a 30\% cap, LC tails within $1.3\times$ of nominal.
Phase asymmetry, not uniform capping, is the foundation for grid-interactive serving.}

\section*{Acknowledgments}
Yueying Li was supported by the NSF under grant CCF-2118709. Yueying Li acknowledges Anvil AI and GPU allocations through allocation CIS230253 from the Advanced Cyberinfrastructure Coordination Ecosystem: Services \& Support (ACCESS) program, which is supported by U.S. National Science Foundation grants \#2138259, \#2138286, \#2138307, \#2137603, and \#2138296. Udit Gupta and Leo Han were supported by NSF Grants CCF-232660 and CCF-2326608, and acknowledge support from Google and Amazon.

\bibliographystyle{IEEEtranS}
\bibliography{main}

\appendix
\subsection{AI Usage}\label{sec:apx-ai}
This section outlines the usage of AI in the generation of this submission.
We used an LLM-based assistant to shorten and polish writing, to provide review-style feedbacks on drafts, and to help with small coding tasks (plotting and build scripts).
All technical content, experiments, and claims were authored and verified by the authors.

\subsection{Proofs and Supplementary Derivations}\label{sec:proofs}

\subsubsection{KKT Derivation for Joint Optimization}\label{sec:proof-dual}
Note in this section, we expand the objective to be involving power in the first term for generality, but it's not necessary. 
The Lagrangian of the joint optimization~\eqref{eq:dual} with multipliers $\mu \ge 0$ (power cap) and $\eta_i \ge 0$ (impact bound per stage-class $i$) is:
\begin{align}\label{eq:lagrangian}
\mathcal{L} = &\;\Delta t \cdot P(\mathbf{k},\mathbf{f}) + \sum_i w_i I_i + \mu\big(P(\mathbf{k},\mathbf{f}) - p_{\max}\big) \nonumber\\
& + \sum_i \eta_i \big(I_i(\capfn_i) - \delta_i\big).
\end{align}

\paragraph{Frequency stationarity.}
Setting $\partial\mathcal{L}/\partial f_i = 0$:
\begin{equation}\label{eq:kkt-f}
\underbrace{(\Delta t + \mu) \cdot \frac{\partial P}{\partial f_i}}_{\text{power cost of raising } f_i}
= \underbrace{-(w_i + \eta_i) \cdot \frac{\partial I_i}{\partial \capfn_i} \cdot \frac{\partial \capfn_i}{\partial f_i}}_{\text{impact benefit of raising } f_i}.
\end{equation}
Under the per-GPU power model $P(f) = c_2 f^3 + c_1 f + c_0$ aggregated over $k_i$ active GPUs, the marginal power cost on the left side is $k_i\,(3 c_2 f_i^2 + c_1)$.
The capacity gradient with respect to frequency is:
\begin{equation}\label{eq:dcap-df}
\frac{\partial \capfn_i}{\partial f_i} = \frac{k_i^{\beta_i} \gamma_i}{f_{\max}},
\end{equation}
which is proportional to the compute-bound fraction $\gamma_i$.

\paragraph{Complementary slackness.}
$\eta_i^\star \ge 0$ and $\eta_i^\star(\delta_i - I_i^\star) = 0$: the multiplier $\eta_i$ is positive only when the impact bound is tight ($I_i = \delta_i$), so stage-classes with SLO headroom ($I_i < \delta_i$) have $\eta_i = 0$ and are degraded first.
\qed

\subsubsection{Deriving the Compute-Bound Fraction \texorpdfstring{$\gamma_i$}{gamma_i}}\label{sec:gamma-derivation}


\paragraph{Per-layer FLOPs and bytes.}
For a standard transformer layer ($d_{\mathrm{ff}}{=}4d$) processing a batch of $B$ decode tokens with context length $S$, the total floating-point operations and memory bytes are:
\begin{align}
F &= \underbrace{24Bd^2}_{\text{MLP + projections}} + \underbrace{4BSd}_{\text{attention}} = 4Bd(6d + S), \label{eq:flops-layer}\\[3pt]
M &= \underbrace{12d^2 \cdot b_w}_{\text{weights}} + \underbrace{2BSd \cdot b_w}_{\text{KV cache}} = 2d(6d + BS)\,b_w, \label{eq:bytes-layer}
\end{align}
where $b_w$ is bytes per parameter (e.g., 2 for bf16).

\paragraph{Arithmetic intensity and $\gamma_i$.}
The overall arithmetic intensity of the layer is:
\begin{equation}\label{eq:ai-layer}
\mathrm{AI} = \frac{F}{M} = \frac{2B(6d + S)}{(6d + BS)\,b_w}.
\end{equation}
The hardware balance point is $I^{\star} = C / W$ (peak FLOP/s over HBM bandwidth).
The layer is fully compute-bound when $\mathrm{AI} \ge I^{\star}$ and fully memory-bound when $\mathrm{AI} \ll I^{\star}$.
The compute-bound fraction is:
\begin{equation}\label{eq:gamma}
\gamma_i = \min\!\Big(1,\;\frac{\mathrm{AI}}{I^{\star}}\Big) = \min\!\bigg(1,\;\frac{2BW(6d + S)}{(6d + BS)\,b_w\,C}\bigg).
\end{equation}

\paragraph{Implications.}
For prefill ($B$ = prompt length, large), $\mathrm{AI}$ grows with $B$ and $\gamma_i \to 1$.
For decode with small $B$, $\mathrm{AI} \approx 2B/b_w \ll I^{\star}$, so $\gamma_i \approx 0$.
Longer context $S$ increases the KV-cache memory term in the denominator, further reducing $\gamma_i$, confirming that decode stages with long reasoning traces are predominantly memory-bound and amenable to frequency reduction with minimal throughput loss.

\subsubsection{Throughput, Power, Latency, and Energy Models}\label{sec:perf-models}

PSOpt's analytical solver relies on the four per-stage models profiled in \S\ref{sec:formulation-solver} and visualized in Figure~\ref{fig:freqscaling4panel}.
All four are fit per-model on A100-80GB at TP-4 across the batch-size breakpoints reported in \S\ref{sec:formulation-solver}.

\paragraph{Latency.}
Per-batch latency is fit as a piecewise-linear function of $f_i$ with a knee at a batch-size-dependent saturation frequency $f^{\star}(\mathrm{BS})$:
\begin{equation}\label{eq:latency-pwl}
L(f_i; \mathrm{BS}) =
\begin{cases}
a_1(\mathrm{BS}) - b_1(\mathrm{BS})\,f_i, & f_i \le f^{\star}(\mathrm{BS}) \\
a_2(\mathrm{BS}) - b_2(\mathrm{BS})\,f_i, & f_i > f^{\star}(\mathrm{BS})
\end{cases}
\end{equation}
with $b_2 \ll b_1$ in the memory-bandwidth-bound regime (low $\gamma_i$, e.g.\ answer decode at small batch), collapsing to a single linear segment when the stage is compute-bound across the full frequency range (high $\gamma_i$, e.g.\ prefill).
This regime-dependent slope is what makes stage-aware DVFS effective: dropping $f_i$ on low-$\gamma_i$ stages beyond $f^{\star}$ cuts power with negligible latency cost.

\paragraph{Throughput.}
Per-instance throughput follows directly from Eq.~\ref{eq:latency-pwl} as $\mathrm{BS} / L(f_i; \mathrm{BS})$.
Figure~\ref{fig:freqscaling4panel} confirms the regime split implied by the two slopes: at small batch size throughput saturates above ${\sim}$810\,MHz, while at large batch size it scales nearly linearly up to 1{,}410\,MHz.

\paragraph{Power.}
Per-GPU power follows the cubic model $P(f) = c_2 f^3 + c_1 f + c_0$ from \S\ref{sec:formulation-solver}, with coefficients $(c_2, c_1, c_0)$ profiled at each batch-size breakpoint; aggregate stage power is $k_i P(f_i)$.
The $R^2$ of the fit exceeds 0.95 across all profiled points.

\paragraph{Energy.}
Per-token energy is $P(f_i)\cdot L(f_i;\mathrm{BS}) / \mathrm{BS}$.
In the compute-bound regime latency drops roughly as $1/f_i$ while power grows as $f_i^3$, so energy-per-token has a minimum at an intermediate frequency; in the memory-bound regime latency is nearly flat beyond $f^{\star}$ so energy-per-token falls monotonically as $f_i$ is reduced, justifying the aggressive frequency reduction PSOpt selects for answer decode.

\subsection{From Static Provisioning to Dynamic Runtime Routing}\label{sec:pta-runtime}

The runtime operates in two layers.
An \emph{allocator} provisions Prefill, Think, and Answer pools once at start-up, fixing the per-stage parallelism configuration and tagging each instance with its home stage.
A \emph{dispatcher} then routes every arriving request across those pools and, under skewed load, re-tags a small number of instances so that the effective pool boundary tracks the offered workload between PSOpt epochs.

\paragraph{Static provisioning.}
Given per-stage pool sizes and parallelism degrees from PSOpt, the allocator partitions the cluster into three contiguous blocks, one per stage, and spawns instances at tensor-parallel granularity within each block.
The result is three sets of stage-tagged instances that define the initial $\mathbf{k}$ in Eq.~\ref{eq:dual}; no further static capacity planning occurs until the next PSOpt epoch.

\paragraph{Dynamic dispatch.}
At arrival, each request is classified as two-phase (Prefill then Answer) or three-phase (Prefill, Think, then Answer) based on whether the model emits a reasoning trace.
For each hop, the dispatcher selects the least-loaded instance in the corresponding pool under a load key $\phi$.
Three variants of $\phi$ are supported -- memory reservation (JSQ), pending-token count (TokenJSQ), and a convex combination of the two (Weighted) -- expressing the same selection rule under different notions of ``load.''
Reservations are updated before the dispatcher returns so that subsequent arrivals observe the new load immediately; KV-cache transfers between selected instances contend on the destination's inbound bandwidth.

\paragraph{Online pool rebalancing.}
Strict stage pools are efficient when the offered mix matches the provisioned split but waste capacity when one stage saturates while another is idle, a common situation for reasoning workloads, whose think-to-answer ratio varies with prompt difficulty.
\sys{} absorbs this mismatch without waiting for the next PSOpt epoch: when a pool has no instance satisfying its load and SLO predicates, the dispatcher \emph{borrows} an underutilized instance from a sibling pool, retags it for the borrowing stage, and places it in a shared ``mixed'' pool visible to both stages.
When a borrowed instance drains its in-flight work, it is released back to its home pool.
This borrow-and-return mechanism turns the static P/T/A boundary into a soft one at the per-request timescale, while PSOpt remains responsible for coarse re-provisioning at epoch boundaries.
Class-visibility filters (a request of class $c$ considers only instances whose reserved load comes from classes at priority $\ge c$) interact with rebalancing without further modification: LC requests see the whole cluster, Flex and BE requests see progressively smaller slices, preserving the SLO-tier ordering from \S\ref{sec:flex-impact}.

Algorithm~\ref{alg:pta-runtime} distills the two-layer runtime: a one-shot \textsc{Provision} step at epoch boundaries and a per-request \textsc{Dispatch} step that invokes \textsc{Rebalance} on demand.

\begin{algorithm}[ht]
\caption{\sys{} static provisioning and dynamic routing.}
\label{alg:pta-runtime}
\small
\begin{algorithmic}[1]
\STATE \textbf{function} \textsc{Provision}($\mathbf{k}$ from PSOpt, per-stage parallelism)
\STATE \quad partition cluster into blocks of size $k_\pi$, one per stage $\pi \in \{\textsf{P},\textsf{T},\textsf{A}\}$
\STATE \quad spawn tensor-parallel instances in each block; tag each by home stage
\STATE \textbf{end function}
\vspace{2pt}
\STATE \textbf{function} \textsc{Dispatch}($R$ of class $c$, load key $\phi$, SLO $\tau_c$)
\STATE \quad $\mathcal{H} \gets$ stage sequence for $R$ \hfill\COMMENT{2-phase or 3-phase}
\FOR{each stage $\pi \in \mathcal{H}$}
  \STATE $i^\star_\pi \gets \arg\min_{i \in \mathrm{pool}(\pi),\,\text{class-visible to } c} \phi(i)$
  \IF{no $i^\star_\pi$ admits $R$ within $\tau_c$}
    \STATE $i^\star_\pi \gets$ \textsc{Rebalance}($\pi$) \hfill\COMMENT{borrow from sibling pool}
  \ENDIF
\ENDFOR
\STATE \quad reserve load on each $i^\star_\pi$ and dispatch $R$ along $\mathcal{H}$
\STATE \quad \textbf{return} $(i^\star_\pi)_{\pi \in \mathcal{H}}$
\STATE \textbf{end function}
\vspace{2pt}
\STATE \textbf{function} \textsc{Rebalance}($\pi$)
\STATE \quad pick underutilized instance $j$ from a sibling pool; retag $j$ to $\pi$
\STATE \quad place $j$ in a mixed pool visible to both stages
\STATE \quad on completion of $j$'s in-flight work, release $j$ to its home stage
\STATE \quad \textbf{return} $j$
\STATE \textbf{end function}
\end{algorithmic}
\end{algorithm}

\subsection{Experimental Setup Details}\label{sec:apx-setup}
Table~\ref{tab:workloads} lists the workloads and their latency targets; Table~\ref{tab:baselines} summarizes the five baselines, all using the same cluster and power-aware routing.

\begin{table}[t]
\centering
\small
\caption{Workloads in evaluation and latency requirements.}
\label{tab:workloads}
\begin{tabular}{lccl}
\toprule
\textbf{Model} & \textbf{TTF(A)T} & \textbf{TTLT/TBT} & \textbf{Dataset} \\
\midrule
CodeLlama-34B         & 72.0s & 115.5s & Magpie     \\
QWen-32B              & 36.0s & 121.0s & S1K        \\
DS-R1-Llama-70B       & 5.0s  & 0.50s  & Azure-Code \\
QWen-14B              & 0.30s & 0.1s   & Azure-Chat \\
\bottomrule
\end{tabular}
\end{table}

\begin{table}[t]
\centering
\scriptsize
\setlength{\tabcolsep}{3pt}
\caption{Baseline configurations.}
\label{tab:baselines}
\begin{tabular}{lllll}
\toprule
Name & Arch & DVFS Policy & Scheduler & Multi-SLO \\
\midrule
\textbf{B1} Uniform     & PD disagg  & Uniform        & Overlap-KV JSQ         & Single class \\
\textbf{B2} POLCA       & Collocated & Priority-aware & Mixed-JSQ              & HP/LP binary \\
\textbf{B3} SplitWise+  & PD disagg  & Stage-aware    & Overlap-KV JSQ         & +Priority \\
\textbf{B4} DynamoLLM+  & Collocated & +Uniform       & Prediction rtr.      & SLO-aware \\
\textbf{B5} SLOs-Serve+ & Collocated & +Uniform       & ProMax                 & LC/BE priority \\
\textbf{\sys{}} & PTA disagg & KKT solver     & Multi-SLO JSQ          & LC/Flex/BE \\
\bottomrule
\end{tabular}
\end{table}

\subsection{Additional Motivation Data}\label{sec:apx-motivation}

\subsubsection{Prediction-Based Routing under Reasoning Workloads}\label{sec:apx-misrouting}

\begin{figure}[t]
  \centering
  \includegraphics[width=\linewidth]{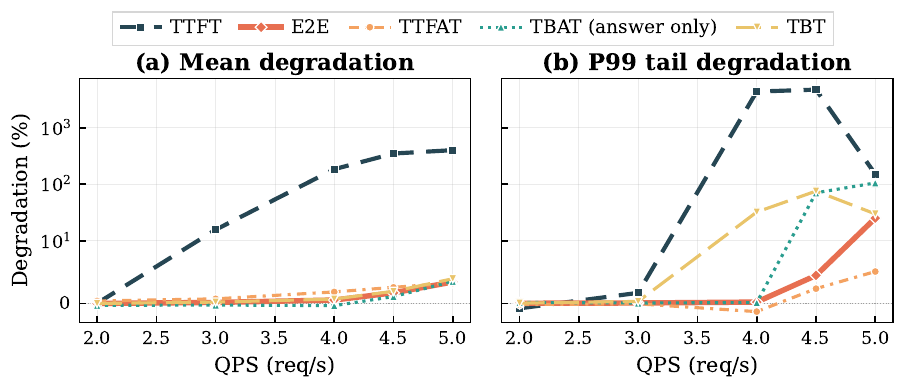}
  \vspace{-0.5em}
  \caption{Latency degradation from prediction-based misrouting under increasing per-instance offered load (128-instance cluster).
    At low load ($\leq$2\,rps per instance) misrouting has negligible impact; near saturation ($\sim$5\,rps per instance), TTFT mean degrades $>$100\% and P99 exceeds 1{,}000\%, while end-to-end and \TTFAT{} remain buffered until saturation.}
  \label{fig:prediction-motivation}
\end{figure}

Figure~\ref{fig:prediction-motivation} details the misrouting cascade summarized in \S\ref{sec:c2}.
A DistBert-style length classifier trained on the reasoning distribution of Figure~\ref{fig:token-distribution} yields a 37.1\% overall misprediction rate with a severe asymmetry: 26.6\% of Medium and 16.3\% of Long requests are misrouted to the Short pool, inflating its load by 25\%; Medium requests are hardest to classify, with only 49.9\% routed correctly even after recalibrating length-class thresholds to equalize class sizes.
At low request rates the impact is negligible, but as the system approaches saturation (the regime a power-cap reduction induces), TTFT rises sharply: mean latency doubles and P99 increases by over an order of magnitude, while per-token metrics stay below 10\%, confirming that the bottleneck is routing-induced queueing rather than per-token compute.
The behavior is not specific to DynamoLLM; it applies to any pre-sized-pool design whenever decode-length variance across request classes is high.

\subsubsection{PTA Benefit across Workload Distributions}\label{sec:apx-workload-sens}

\begin{figure}[t]
  \centering
  \includegraphics[width=\linewidth]{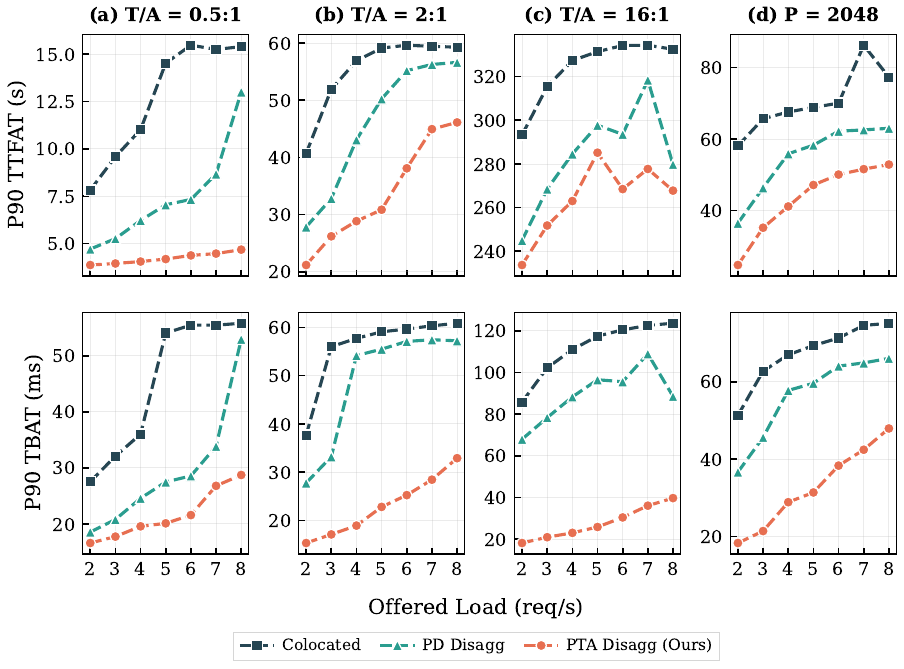}
  \vspace{-0.5em}
  \caption{Robustness of the PTA benefit across workload distributions. Panels (a)--(c) increase the T/A ratio from 0.5:1 to 16:1 (average prefill 512); panel (d) uses average prefill 2048. The advantage grows with T/A ratio (49\% TBAT reduction at 0.5:1 up to 71\% at 16:1 at high load) because longer thinking chains cause more severe batch interference under PD and colocated serving, which PTA eliminates by isolating answer decode onto dedicated workers.}
  \label{fig:workload-sensitivity}
\end{figure}

Figure~\ref{fig:workload-sensitivity} sweeps the think-to-answer ratio and prefill length behind the Observation~3 numbers (\S\ref{sec:c2}), showing the PD-vs-PTA gap across the workload-distribution space.

\subsection{Additional Experiments}\label{sec:apx-planned}

\subsubsection{Per-Request Transient Behavior}\label{sec:apx-transient}

\begin{figure}[t]
  \centering
  \includegraphics[width=\linewidth]{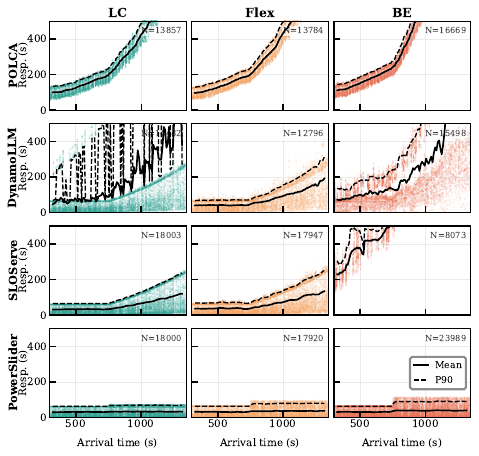}
  \vspace{-0.5em}
  \caption{Per-request response times for LC, Flex, and BE classes as the power cap steps from 90\% to 60\% of nominal at $t=750$\,s on a 512-GPU cluster. \sys{} completes the most requests across all classes with stable response times; baselines either shed BE traffic or accumulate unbounded queueing.}
  \label{fig:timeline-cap90-60}
\end{figure}

Figure~\ref{fig:timeline-cap90-60} shows the per-request scatter behind the Q5 summary (\S\ref{sec:eval-convergence}): response times for every completed request in all three classes, for \sys{} and the three strongest baselines, across the cap step.

\begin{figure}[t]
    \centering
    \includegraphics[width=0.7\columnwidth]{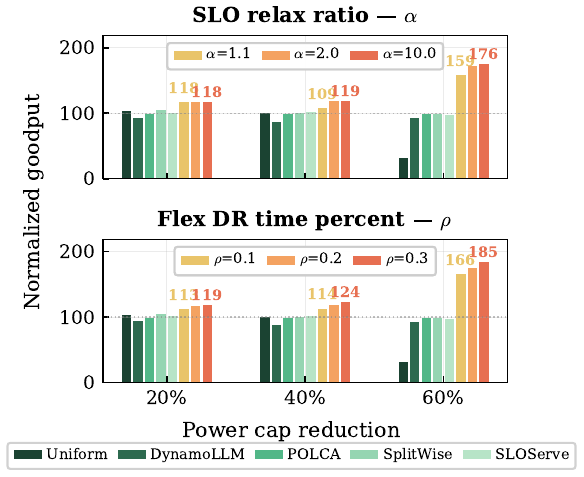}
        \vspace{-0.5em}
    \caption{Sensitivity of \sys{}'s Flex SLO parameters on LC goodput, normalized to POLCA.
    \textbf{Top:} varying the relaxation ratio $\alpha$ with $\rho = 0.2$.
    \textbf{Bottom:} varying the violation-budget fraction $\rho$ with $\alpha = 3.0$.
    At a 60\% power-cap reduction, \sys{} delivers up to $1.85\times$ the goodput of POLCA and remains robust over a wide range of $\alpha$ and $\rho$.
    \rev{The contract that funds \sys{}'s power headroom does not need careful tuning to deliver it.}}
    \label{fig:flex-sensitivity}
\end{figure}

\begin{figure*}[t]
  \centering
  \includegraphics[width=0.78\textwidth]{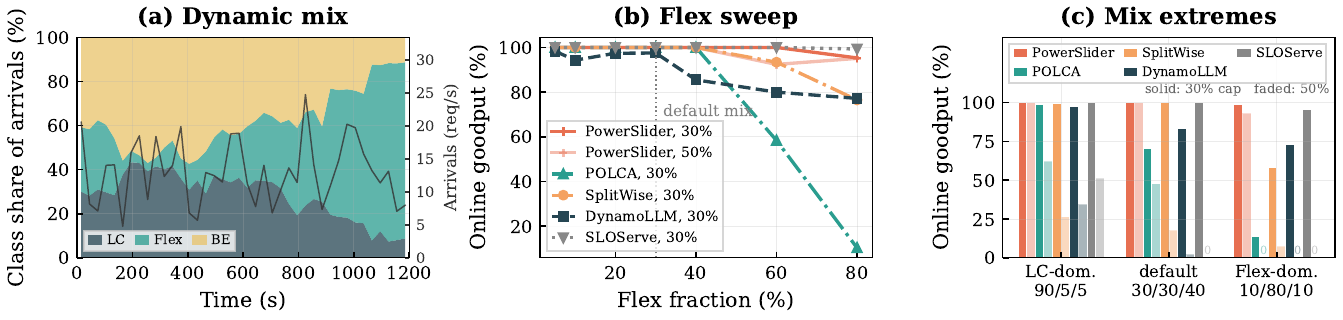}
  \caption{\textbf{Workload-mix sensitivity under a dynamic user mix (measured).} (a) The LC/Flex/BE arrival mix drifts from 5\% to 80\% Flex \emph{within a single run} (QPS\,14, CoV\,7.5 bursty arrivals, 64 GPUs); the gray line is the arrival rate. (b) Online goodput per mix segment at a 30\% power-cap reduction (\sys{} also shown at 50\%): \sys{} holds 95--100\% goodput across the entire drift (95.3\% at 80\% Flex, shedding BE admissions when the cap makes full service infeasible), while POLCA collapses to 10\% at 80\% Flex. (c) Mix extremes -- LC-dominant (90/5/5), default (30/30/40), Flex-dominant (10/80/10) -- at 30\% (solid) and 50\% (faded) reductions.}
  \label{fig:adv-mix}
\end{figure*}

\begin{figure*}[t]
  \centering
  \includegraphics[width=0.78\textwidth]{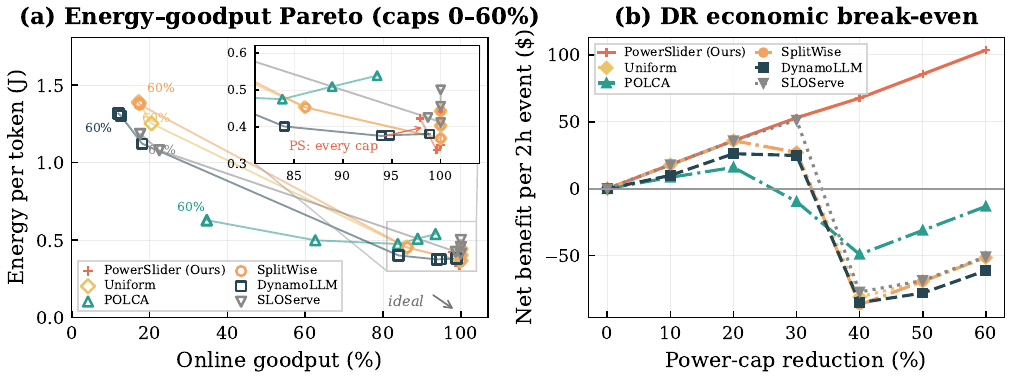}
  \caption{\textbf{Energy--goodput Pareto and economic break-even under power caps (measured).} \textbf{(a)} Cluster energy per generated token vs.\ online goodput, one point per system and cap depth (0--60\%; ideal is bottom-right). \sys{} alone holds the ideal corner at every depth (0.34--0.44\,J/token at 97.9--100\% goodput; inset), while at a 60\% cap the baselines pay $\approx$3$\times$ the energy per \emph{useful} token at 12--23\% goodput. \textbf{(b)} Net benefit per 2\,h DR event vs.\ cap depth; \sys{} is net-positive at every depth where the baselines turn negative (\S\ref{sec:eval-generality}).}
  \label{fig:adv-energy-token}
  \label{fig:adv-econ}
\end{figure*}

This appendix collects additional results referenced from \S\ref{sec:eval}, workload-mix sensitivity (Figure~\ref{fig:adv-mix}), economic break-even and the energy--goodput Pareto (Figure~\ref{fig:adv-econ}), and grid-program feasibility (Table~\ref{tab:adv-grid-programs}).


\textbf{Workload-mix sensitivity.}
Figure~\ref{fig:adv-mix} quantifies how much of the Q1 advantage depends on the traffic mix, using a mix that drifts within one run rather than separate static runs. \sys{} is essentially mix-insensitive: it sustains 95--100\% online goodput over the whole 5\%$\to$80\% Flex drift at a 30\% power-cap reduction (92--100\% at 50\%), shedding BE admissions rather than violating online SLOs when the cap makes full service infeasible, and 93--100\% at every mix extreme. Baselines degrade as the Flex share grows because they lack an interface to exploit contractual slack: at a 30\% reduction, POLCA falls from 100\% at $\leq$40\% Flex to 58\% at 60\% and 10\% at 80\% Flex; at the Flex-dominant extreme (10/80/10) it retains 14\% (30\% reduction) and 0\% (50\%) vs.\ \sys{}'s 93--99\%.



\paragraph{Generalization beyond A100.}
The cubic DVFS power model fitted on GH200 ($R^2{=}0.98$) also fits public H100 power: the same cubic form holds with refitted coefficients. Porting \sys{} to a new hardware type therefore requires only that per-hardware refit of the power cubic and throughput points; the simulator already carries an H100 profile scaled from the measured A100 coefficients, with the wider 210--1980\,MHz frequency range and 700\,W TDP. We expect the wider DVFS range to increase goodput retention under deep caps; a full H100 evaluation is left to future work.

\textbf{Energy per useful token.}
Figure~\ref{fig:adv-energy-token}(a) recasts the goodput results in energy terms. At every cap depth \sys{}'s energy per generated token stays within 0.34--0.44\,J: at a 60\% cap reduction it still serves 97.9\% of online load at 0.42\,J/token, equal-or-better than its own uncapped 0.44\,J, because stage-aware DVFS slows memory-bound pools where throughput is nearly frequency-insensitive. The baselines' energy per token degrades $\approx$3$\times$ at deep caps not because they draw more power but because most of that energy goes to requests that miss their SLOs or never complete; POLCA avoids the worst of this by shedding load, plateauing at 0.63\,J/token with 35\% goodput. Joules per useful token is also the quantity operational-carbon models consume.

\textbf{Economic break-even.}
Figure~\ref{fig:adv-econ}(b) converts measured goodput into dollars: DR enrollment pays for curtailed energy, while cap-induced goodput loss forfeits serving revenue.
\rev{We assume scarcity-event compensation of \$1.75/kWh of committed curtailment (emergency-response-service pricing; economic-DR prices outside scarcity events are orders of magnitude lower) and \$2.50/GPU-hr serving revenue. Because every system receives the same payment at a given depth, the ranking below is price-independent; the absolute break-even depths scale with the curtailment-to-revenue price ratio.}
Under these prices, every baseline's revenue loss overtakes the DR payment at moderate depths -- POLCA first at $\approx$27\%, Uniform at $\approx$33\%, the remaining baselines at 32--34\% -- consistent with operators' current reluctance to enroll inference clusters in DR programs. Because \sys{} holds online goodput at 97.9--100\% at every depth, its revenue loss never exceeds \$4 per event and net benefit grows monotonically with depth, reaching \$103 per 2\,h event at a 60\% reduction \rev{under scarcity-priced events}.

\paragraph{Multi-tier grid-program participation.}
Table~\ref{tab:adv-grid-programs} assesses which formal grid programs \sys{}'s actuation latencies can serve. Two response paths exist: the DVFS-only band (up to a $\sim$15--20\% power-cap reduction) actuates in under a second (7.7\,ms solve $+$ 100\,ms DVFS interval), fast enough for AGC-driven frequency regulation; deeper responses require one 5\,min drain-bounded reallocation epoch, comfortably inside spinning-reserve (10\,min) and emergency-DR (30\,min) windows. Because the two bands actuate independently, they can be \emph{stacked}: enroll the DVFS band in regulation while committing the reallocation band to reserves, monetizing both fast and deep flexibility from the same cluster.

\begin{table}[t]
\centering
\caption{Grid-program feasibility given \sys{}'s actuation paths (7.7\,ms solve; $<$1\,s DVFS band; 5\,min reallocation epoch).}
\label{tab:adv-grid-programs}
\scriptsize
\setlength{\tabcolsep}{3pt}
\begin{tabular}{@{}llll@{}}
\toprule
\textbf{Program} & \textbf{Resp.\ req.} & \textbf{\sys{} path} & \textbf{Feasible} \\
\midrule
Freq.\ regulation (AGC) & seconds & DVFS band ($<$1\,s) & \cmark\ ($\leq$20\%) \\
Spinning reserve & 10\,min & reallocation epoch & \cmark\ (full depth) \\
Emergency DR & 30\,min & reallocation epoch & \cmark\ (full depth) \\
Day-ahead / economic & hours & capacity planning & \cmark \\
Stacked (reg.\ $+$ reserve) & both & both bands & \cmark \\
\bottomrule
\end{tabular}
\end{table}

\end{document}